\documentclass[lettersize,journal]{IEEEtran}
\usepackage{amsmath,amsfonts}
\usepackage{algorithmic}
\usepackage{algorithm}
\usepackage{array}
\usepackage[caption=false,font=normalsize,labelfont=sf,textfont=sf]{subfig}
\usepackage{textcomp}
\usepackage{stfloats}
\usepackage{url}
\usepackage{verbatim}
\usepackage{graphicx}
\usepackage{cite}
\usepackage[english]{babel}
\usepackage{graphicx}
\usepackage{framed}
\usepackage[normalem]{ulem}
\usepackage{amsmath}
\usepackage{amsthm}
\usepackage{amssymb}
\usepackage{amsfonts}
\usepackage{enumerate}
\usepackage[utf8]{inputenc}
\usepackage{euscript}
\usepackage{xcolor}
\usepackage{cuted}
\usepackage{multirow}
\usepackage{booktabs}
\usepackage{xfrac} 
\usepackage{balance}

\usepackage[many]{tcolorbox}
\newtheorem{assumption}{Assumption}
\tcolorboxenvironment{assumption}{
  colback=gray!2!white,
  boxrule=0.15pt,
  boxsep=0.5pt,
  left=2pt,right=2pt,top=2pt,bottom=2pt,
  oversize=0pt,
  rounded corners,
  before skip=\topsep,
  after skip=\topsep,
}
\newtheorem{remark}{Remark}
\tcolorboxenvironment{remark}{
  colback=gray!2!white,
  boxrule=0.15pt,
  boxsep=0.5pt,
  left=2pt,right=2pt,top=2pt,bottom=2pt,
  oversize=2pt,
  rounded corners,
  before skip=\topsep,
  after skip=\topsep,
}
\newtheorem{theorem}{Theorem}
\tcolorboxenvironment{theorem}{
  colback=gray!2!white,
  boxrule=0.15pt,
  boxsep=0.5pt,
  left=2pt,right=2pt,top=2pt,bottom=2pt,
  oversize=2pt,
  rounded corners,
  before skip=\topsep,
  after skip=\topsep,
}
\newtheorem{definition}{Definition}
\tcolorboxenvironment{definition}{
  colback=gray!2!white,
  boxrule=0.15pt,
  boxsep=0.5pt,
  left=2pt,right=2pt,top=2pt,bottom=2pt,
  oversize=2pt,
  rounded corners,
  before skip=\topsep,
  after skip=\topsep,
}
\newtheorem{corollary}{Corollary}
\tcolorboxenvironment{corollary}{
  colback=gray!2!white,
  boxrule=0.15pt,
  boxsep=0.5pt,
  left=2pt,right=2pt,top=2pt,bottom=2pt,
  oversize=2pt,
  rounded corners,
  before skip=\topsep,
  after skip=\topsep,
}

\newcommand{\myargmax}[1]{\underset{{#1}}{\text{argmax}}}

\newcommand{\minimize}[1]{\underset{{#1}}{\text{minimize}}}
\newcommand{\maximize}[1]{\underset{{#1}}{\text{maximize}}}
\newcommand{\st}{\text{subject to}}

\usepackage{mathtools}
\newcommand{\norm}[1]{\left\|#1 \right\|}

\newcommand\mydots{\hbox to 0.75em{.\hss.\hss.}}
\newcommand{\mb}[1]{\mathbf{#1}}
\newcommand{\mbg}[1]{\boldsymbol{#1}}
\newcommand\smallmath[2]{#1{\raisebox{\dimexpr \fontdimen 22 \textfont 2
      - \fontdimen 22 \scriptscriptfont 2 \relax}{$\scriptstyle #2$}}}
\newcommand\smallotimes{\smallmath\mathbin\otimes}
\usepackage{stackengine,graphicx}
\newcommand\No[1][.13ex]{%
  \setbox0=\hbox{\scalebox{.7}{o}}%
  \setbox2=\hbox{N}%
  N\kern-.05em\stackengine{\dimexpr\ht0-\ht2+#1}{\belowbaseline[-\ht2]{\copy0}}%
    {\rule[-.13ex]{.7\wd0}{.13ex}}%
    {U}{c}{F}{F}{L}%
}

\newcommand{\blue}[1]{\textcolor{black}{#1}}

\begin{document}

\begingroup
\allowdisplaybreaks

\title{Regression Equilibrium in Electricity Markets}

\author{Vladimir Dvorkin,~\IEEEmembership{Member,~IEEE}
\thanks{V. Dvorkin is with the Department of Electrical Engineering and Computer Science, University of Michigan, Ann Arbor, MI. E-mail: dvorkin@umich.edu}}

\markboth{IEEE Transactions on Energy Markets, Policy and Regulation}%
{Shell \MakeLowercase{\textit{et al.}}: A Sample Article Using IEEEtran.cls for IEEE Journals}

\maketitle

\begin{abstract}
\blue{In two-stage electricity markets, renewable power producers enter the day-ahead market with a forecast of future power generation and then reconcile any forecast deviation in the real-time market at a penalty. The choice of the forecast model is thus an important strategy decision for renewable power producers as it affects financial performance. In electricity markets with large shares of renewable generation, the choice of the forecast model impacts not only individual performance but also outcomes for other producers. In this paper, we argue for the existence of a competitive regression equilibrium in two-stage electricity markets in terms of the parameters of private forecast models informing the participation strategies of renewable power producers. In our model, renewables optimize the forecast against the day-ahead and real-time prices, thereby maximizing the average profits across the day-ahead and real-time markets. By doing so, they also implicitly enhance the temporal cost coordination of day-ahead and real-time markets.} We base the equilibrium analysis on the theory of variational inequalities, providing results on the existence and uniqueness of regression equilibrium in energy-only markets. We also devise two methods to compute regression equilibrium: centralized optimization and a decentralized ADMM-based algorithm.
\end{abstract}


\section*{Selected Nomenclature}
\subsection{Optimization Parameters}
\begin{IEEEdescription}[\IEEEusemathlabelsep\IEEEsetlabelwidth{xxxx}]
\item[$\mb{c}/\mb{C}$] $1^{\text{st}}/2^{\text{nd}}$-order generation cost coefficients 
\item[$\mb{c}^{+}/\mb{c}^{-}$] Vectors of upward$/$downward regulation cost 
\item[$\mb{s}/\mb{S}$] $1^{\text{st}}/2^{\text{nd}}$-order load shedding cost coefficients 
\item[$\widehat{\mb{w}}/\mb{w}$] Vectors of wind power forecast$/$realization
\item[$\mb{d}$] Vector of nodal electricity demands
\item[$\mb{F}$] Matrix of the power transfer distribution factors
\item[$\overline{\mb{f}}$] Vector of maximum transmission line capacities 
\item[$\overline{\mb{p}}/\underline{\mb{p}}$] Vectors of maximum$/$minimum generation capacity 
\item[$\mbg{\Gamma}$] Diagonal matrix of private revenue vs. loss weights
\item[$\varrho$] Step size of the ADMM algorithm 
\item[$\mbg{\varphi}$] Vector of wind power generation features 
\item[$\tau$] Scalar for Lasso regularization 
\end{IEEEdescription}

\subsection{Optimization Variables}
\begin{IEEEdescription}[\IEEEusemathlabelsep\IEEEsetlabelwidth{xxxx}]
\item[$\mb{p}$] Vector of day-ahead generator dispatch
\item[$\mb{r}^{+}/\mb{r}^{\blue{-}}$] Vectors of real-time upward$/$downward regulation 
\item[$\mb{g}$] Stack of  generator decisions, i.e., $\mb{g}=(\mb{p},\mb{r}^{+},\mb{r}^{-})$
\item[$\mb{G}$] Generation profile, i.e.,  $\mb{G}=(\mb{g}_{1},\dots,\mb{g}_{n})$
\item[$\mbg{\ell}$] Vector of real-time load shedding
\item[$\mb{L}$] Load shedding profile, i.e.,  $\mb{L}=(\mbg{\ell}_{1},\dots,\mbg{\ell}_{n})$
\item[$\mbg{\theta}$] Vector of private regression parameters
\item[$\mbg{\Theta}$] Regression profile, i.e.,  $\mbg{\Theta}=(\mbg{\theta}_{1},\dots,\mbg{\theta}_{b})$
\item[$\mbg{\lambda}_{1}/\mbg{\lambda}_{2}$] Day-ahead$/$real-time location marginal prices
\end{IEEEdescription}

\subsection{Special Notation} 
Boldface lowercase$/$uppercase letters denote column vectors$/$matrices. Norm $\norm{\mb{x}}_{\mb{C}}^{2}=\mb{x}^{\top}\mb{C}\mb{x}$. $f(x\;\!|\;\!y)$ is a function of $x$ with parameter $y$. Notation $\mb{x}\perp \mb{y}$ means ``orthogonal'' ($\mb{x}\perp \mb{y} \Leftrightarrow \mb{x}^{\top}\mb{y}=\mb{0}$). For a set $\mb{x}=(x_{1},\dots,x_{i},\dots,x_{n})$, $\mb{x}_{-i}$ collects all elements in $\mb{x}$ except that at position $i$.  Operator $[x]_{+}$ is the projection of $x$ onto non-negative orthant, $\smallotimes$ is the Kronecker product, $\mb{I}_{n}$ is $n\times n$ identity matrix, and vector $\mb{1}$ $(\mb{0})$ is a vector of ones (zeros). 

\section{Introduction}\label{sec:intro}

\IEEEPARstart{W}{holesale} electricity markets are designed to maximize social welfare while maintaining the power supply and demand in balance at all times. However, with the increasing integration of stochastic energy resources,  markets could struggle to maintain this balance in a welfare-maximizing manner. One reason is the lack of coordination between the day-ahead market, which is cleared based on forecasts of renewable power generation, and the real-time market, which offsets any forecast deviation, typically at higher costs to the system. To improve their coordination, several mechanisms have proposed to integrate probabilistic information on renewable generation into day-ahead markets, e.g., using scenario-based \cite{pritchard2010single,morales2012pricing,zavala2017stochastic,ndrio2021pricing} or chance-constrained \cite{dvorkin2019chance,mieth2020risk,ratha2023moving} programming, thereby reducing the spillover from day-ahead markets with imperfect forecasts to real-time markets. However, their adoption means altering the existing market-clearing model and seeking a multifaceted consensus of stakeholders (e.g., stochastic producers) on uncertainty parameters—two currently unresolved challenges in practice.
 
There is potential for more seamless market coordination in enhancing renewable power forecasts. In real-time markets with typically asymmetric regulation costs, over- and under-predictions of stochastic generation are penalized differently, providing wind power producers with incentives to bias their forecasts to target cheaper regulation \cite{bathurst2002trading,pinson2007trading,pinson2023distributionally}. Perusing cheaper regulation, renewable power producers implicitly achieve the least-cost dispatch across the day-ahead and real-time markets \cite{morales2014electricity}. While \cite{bathurst2002trading,pinson2007trading,pinson2023distributionally,morales2014electricity} focus on tuning a point or probabilistic forecasts regardless of the underlying prediction model, the work in \cite{zhang2023deriving,zhang2023value,wahdany2023more,dvorkin2023price} develops algorithms to train machine learning (ML) models which directly map features into decision-focused forecasts. 

As electricity markets are gradually getting dominated by renewables, it is expected that more producers will optimize forecasts in a decision-focused manner. However, unlike in many other domains for ML applications, in power systems the agent strategies are coupled via shared grid infrastructure (physical coupling) and electricity market clearing (economic coupling). The individual choices of prediction models thus lead to ripple effects on the whole system and markets. The work in \cite{dvorkin2023price} indeed shows how the choice of prediction models affects locational marginal prices \blue{(LMPs)} across the entire grid, with average pricing errors—relative to ground truth prices—between \$0.62 and \$11.15 per MWh. 

\subsubsection*{Contributions} This paper studies how stochastic producers can cohesively \blue{extract more revenues from their forecast models in competitive electricity markets while implicitly enhancing the market efficiency across the day-ahead and real-time markets}. In the setting of interest, the ML model selection by each producer affects the revenues of other producers via competitive price formation in the day-ahead and real-time markets. Since private forecast models may be complementary or conflicting to the revenue-seeking objectives of other participants, we use a game-theoretic lens. We establish the existence of an equilibrium among ML models in electricity markets, in which no market participant benefits by unilaterally deviating from their equilibrium model parameters. Importantly, the study reveals the role of this equilibrium in coordinating day-ahead and real-time markets and maximizing the expected market welfare. More specifically,
\begin{itemize}
    \item[1.] We argue for the existence of a Nash equilibrium among private forecast models that inform trading strategies of renewable power producers in electricity markets. We develop an equilibrium model in which profit-seeking \blue{renewables optimize forecast models against day-ahead and real-time prices, eventually} converging to a profit-maximizing state with no incentives to unilaterally change their forecast models. \blue{Importantly, we discovered this equilibrium within the existing two-stage market design.}
    \item[2.] We prove that the equilibrium regression profile—the ensemble of private forecast models at equilibrium—not only benefits individual producers but also leads to a socially optimal solution minimizing the \blue{regularized} dispatch cost. Towards this result, we reformulate the equilibrium problem as a variational inequality (VI) problem and apply the well-established theory of VIs \cite{facchinei2003finite}.
    \item[3.] We develop two methods to compute the regression equilibrium: an equivalent centralized optimization problem, and a decentralized algorithm based on the alternating direction method of multipliers (ADMM) \cite{boyd2011distributed}. 
\end{itemize}

\blue{Our study on the standard IEEE 24-RTS system revealed the important role of network effects. Specifically, the equilibrium regression models are significantly influenced by day-ahead and real-time LMPs, rather than solely by the physical processes underlying the data. As demonstrated in Sec. IV (Fig. \ref{fig:prediction}), it is also optimal for the system to consistently under- or over-predict wind power generation depending on the position of wind farms in the grid.} 

\blue{While the equilibrium model considers wind power producers as the sole users of ML, the private and social benefits are evident even within this narrow setting. As ML applications in electricity markets continue to grow, it is reasonable to include more ML users, such as conventional generators, demand, and system operators, in the model. We leave this for future work.}
\subsubsection*{Related Work} The barriers for implementing the stochastic market-clearing models in \cite{pritchard2010single, morales2012pricing, zavala2017stochastic, ndrio2021pricing, dvorkin2019chance, mieth2020risk, ratha2023moving} motivated several strategies to coordinate the day-ahead and real-time markets within existing, deterministic market-clearing models. The strategy in \cite{morales2014electricity} computes the optimal day-ahead wind power offering by anticipating the cost of real-time re-dispatch across a fixed set of scenarios. This strategy has shown a cost-saving potential of $\approx25$\% on average in the New York ISO system \cite{zhao2023uncertainty}. Similarly, tuning operating reserve \blue{\cite{mays2021quasi,dvorkin2018setting}}, ramping \cite{wang2014flexible} and transmission capacity \cite{jensen2017cost} requirements minimizes the spillovers from day-ahead to real-time markets. More recently, \cite{mieth2023prescribed} proposed a data-driven strategy via cost-optimal optimization of uncertainty sets for computing the ramping and transmission requirements in day-ahead markets. 

Importantly, the work in \blue{\cite{morales2014electricity,zhao2023uncertainty,mays2021quasi,dvorkin2018setting,wang2014flexible,jensen2017cost,mieth2023prescribed}} offers enhancements from the system operator's perspective. In this paper, we establish that the market coordination can be improved without operator's supervision, solely when profit-seeking stochastic producers adhere to their equilibrium ML models.

Decision-focused learning for improving the value of renewable power predictions has gained momentum thanks to the work in \cite{carriere2019integrated,garcia2021application,zhang2023value,zhang2023deriving,zhang2022cost,wahdany2023more,mieth2023prescribed,dvorkin2023price}. The mainstream approaches rely on either end-to-end learning---which leverages feedback from the downstream dispatch optimization using \blue{bilevel optimization \cite{garcia2021application}} or modified variants of the gradient descent \cite{zhang2023value,wahdany2023more,dvorkin2023price,mieth2023prescribed}---or system-optimal loss function, subsequently used in standard training algorithms  \cite{zhang2023deriving,zhang2022cost}. \blue{The virtue of these methods is that they enhance market efficiency without introducing changes to the existing market design. Moreover, decision-focused learning provides the means for implicit risk management. In words of  \cite{garcia2021application}: ``the [decision-focused learning] framework finds support in current industry practices, where \textit{ad hoc} procedures are implemented to bias load forecasts to empirically reduce risks.'' Following the same principles, the regression equilibrium provides the means for implicit risk management and can be achieved in existing markets without altering the market design.}

The works above have taken a system operator's perspective, and how to achieve system-optimal learning among independent market participants remains an open question. Partially, this question was addressed in \cite{dvorkin2023agent}, where two agents, representing two coupled infrastructure systems, reach consensus on the {\it single} ML model for their cost-optimal coordination. In this work, we develop algorithms to coordinate a {\it population} of private ML learning models towards system-optimal results.

\blue{Finally, this work complements studies of competitive electricity market equilibrium and the ability of marginal pricing to ensure socially optimal outcomes. Marginal pricing has been proven to ensure sufficient regulation capacity in real-time markets \cite{jiang2024optimal}. The work in \cite{pritchard2010single, morales2012pricing, zavala2017stochastic, dvorkin2019electricity, dvorkin2019chance} established the existence of marginal prices supporting stochastic equilibrium, which implicitly minimizes the expected cost across the day-ahead and real-time markets. In conjunction with risk-hedging instruments, the ability of marginal prices to ensure socially optimal outcomes was established in \cite{gerard2018risk, mays2023financial}. Marginal prices have been shown to provide virtual bidders with incentives to arbitrage between day-ahead and real-time markets, thereby implicitly leading to socially optimal outcomes \cite{hogan2016virtual, kazempour2017value}. The proposed equilibrium model complements this line of work by discovering the power of marginal pricing in energy-only markets to provide incentives for wind power producers to optimize the ML models against day-ahead and real-time prices (Theorem \ref{th:ne_existence_uniqness}), thereby implicitly minimizing the social cost of electricity (Corollary \ref{corollary}).}

\section{Electricity Market Preliminaries}

Here, we present market settings in Sec. \ref{subsec:setting} and then decision-focused wind power forecasting in Sec. \ref{subsec:forecasting}. 

\subsection{Market Setting}\label{subsec:setting}

We consider a conventional market settlement with day-ahead and real-time stages. The day-ahead market is cleared on day $d-1$ to compute the optimal contracts for power delivery on day $d$, usually on an hourly basis. The real-time market then settles any supply or demand deviation from the hourly day-ahead contracts on day $d$. Similar to \cite{morales2014electricity,zhang2023deriving}, we assume: 
\begin{assumption}\label{ass:1}
    \blue{We assume an energy-only electricity market settlement, where the generation units are solely remunerated for their variable costs of generation \cite{pritchard2010single,morales2014electricity}.}
\end{assumption}
The energy-only settlement remunerates generation units for variable costs associated with the day-ahead dispatch and the real-time regulation; reserve, auxiliary service and unit commitment decisions are outside the scope of the energy-only settlement. \blue{Disregarding unit commitment constraints allows for a {\it convex} market setting, where the effect of machine learning errors on equilibrium is more evident and isolated from impacts of integrality constraints, which are non-convex.} To ease the presentation, we allow each bus to host no more than one conventional generator, one wind power producer and one load. The loads are modeled as \blue{inelastic} (fixed), price-taking demands in the day-ahead market \blue{that are willing to shed a part of their consumption in the real-time market but at a very high cost for the system}. The market welfare maximization problem thus reduces to minimizing the social cost of electricity. We also disregard the ramping of flexible generating units and thus drop the hour index in the presentation. The last three assumptions are also in the spirit of \cite{morales2014electricity,zhang2023deriving} and are not restrictive for the main result, meaning that the equilibrium model can be adapted to relax these assumptions in a straightforward way. \blue{We model power flows using a DC power flow equations} incorporating the matrix of power transfer distribution factors that converts the net bus power injections into power flows, thus avoiding explicit voltage modeling and reducing the set of equilibrium variables to power generation. \blue{It will be revealed later that the network constraints have a significant impact on regression equilibrium.} 

The day-ahead market-clearing then takes the form: 
\begin{subequations}\label{prob:day_ahead_mc}
    \begin{align}
        \minimize{\mb{p}}\quad&\norm{\mb{p}}_{\mb{C}}^{2}+\mb{c}^{\top}\mb{p}\label{day_ahead_mc_obj}\\
        \st\quad&\mb{1}^{\top}(\mb{p} +\widehat{\mb{w}} - \mb{d})=0,\quad\colon\mu_{1}\label{day_ahead_mc_pb}\\
        &|\mb{F}\;\!\;\!(\mb{p} + \widehat{\mb{w}} - \mb{d})| \leqslant \mb{\overline{f}},\quad\!\colon\underline{\mbg{\kappa}}_{1},\overline{\mbg{\kappa}}_{1}\label{day_ahead_mc_flow}\\
        &\mb{\underline{p}}\leqslant\mb{p}\leqslant\mb{\overline{p}}.\label{day_ahead_mc_gen}
    \end{align}
\end{subequations}
Given the vector of electricity demand $\mb{d}$ and the forecast of wind power generation $\widehat{\mb{w}}$, the goal here is to find the optimal day-ahead dispatch $\mb{p}^{\star}$ of conventional generation, that minimizes generation costs $\eqref{day_ahead_mc_obj}$ while satisfying power grid constraints  \eqref{day_ahead_mc_pb} through \eqref{day_ahead_mc_gen}. \blue{The variable costs of generation include the fuel component only, so they can be represented by a quadratic function \cite[\S11.7]{bergen2009power}}. \blue{The first constraint \eqref{day_ahead_mc_pb}  is a network-wide power balance condition, which requires the total demand $\mb{1}^{\top}\mb{d}$ to be met by the total conventional generation and aggregated wind power forecast  $\mb{1}^{\top}(\mb{p} + \widehat{\mb{w}}).$ Constraint \eqref{day_ahead_mc_flow} enforces the power flow limits. Here, the power flows are computed as the product of the PTDF matrix $\mb{F}$ and the net nodal injection $\mb{p} + \widehat{\mb{w}} - \mb{d}$, whose magnitude is then bounded by the maximum line capacity $\mb{\overline{f}}$. The last constraint \eqref{day_ahead_mc_gen} requires that the day-ahead power dispatch remains within technical generation limits.}

Given \blue{the day-ahead dispatch $\mb{p}^{\star}$ and} the actual realization of wind power $\mb{w}$, the real-time market optimizes:
\begin{subequations}\label{prob:real_time_mc}
    \begin{align}
        \minimize{\mb{r}^{+},\mb{r}^{-},\mbg{\ell}}\quad&\norm{\mb{p}^{\star}+\mb{r}^{+}-\mb{r}^{-}}_{\mb{C}}^{2}+\mb{c}^{+\top}\mb{r}^{+}-\mb{c}^{-\top}\mb{r}^{-}\nonumber\\
        &+\norm{\mbg{\ell}}_{\mb{S}}^2+ \mb{s}^{\top}\mbg{\ell}\label{real_time_mc_obj}\\
    \st\quad
    &\mb{1}^{\top}(\mb{r}^{+}-\mb{r}^{-}+\mb{w}-\widehat{\mb{w}}+\mbg{\ell}) = 0,\;\;\;\!\;\;\;\;\;\!\;\!\colon\mu_{2}\label{real_time_mc_bal}\\
    &|\mb{F}(\mb{p}^{\star} + \widehat{\mb{w}} - \mb{d}) \nonumber\\
    &\!+\mb{F}(\mb{r}^{+}-\mb{r}^{-}+\mb{w}-\widehat{\mb{w}}+\mbg{\ell}) | \leqslant \mb{\overline{f}},\!\colon\!\underline{\mbg{\kappa}}_{2},\overline{\mbg{\kappa}}_{2}\label{real_time_mc_flow}\\
    &\mb{\underline{p}}\leqslant\mb{p}^{\star}+\mb{r}^{+}-\mb{r}^{-}\leqslant\mb{\overline{p}},\label{real_time_mc_gen_0}\\
    &\mb{0}\leqslant\mb{r}^{+}\leqslant\mb{\overline{p}}-\mb{p}^{\star},\label{real_time_mc_gen_1}\\
    &\mb{0}\leqslant\mb{r}^{-}\leqslant\mb{p}^{\star}-\mb{\underline{p}}\label{real_time_mc_gen_2},\\
    &\mb{0}\leqslant\mbg{\ell}\leqslant\mb{d}\label{real_time_load_shed_lims},
    \end{align}
\end{subequations}
that minimizes the cost of re-dispatch \eqref{real_time_mc_obj} by selecting the optimal upward and downward regulation of generating units, $\mb{r}^{+}$ and $\mb{r}^{-}$, respectively, and load shedding $\mbg{\ell}$ in extreme cases. An important feature of the energy-only settlement is the cost order $\mb{c}^{+}\!>\!\mb{c}\!>\!\mb{c}^{-}$, which ensures that either positive or negative deviations from the forecast always result in additional costs to the power system. Equation \eqref{real_time_mc_bal} requires the balance of regulation actions w.r.t. forecast errors, and inequality \eqref{real_time_mc_flow} guarantees that the corresponding adjustment of day-ahead power flows remains within the line limits. Finally, the real-time regulation is bounded by the day-ahead contracts and technical limits via constraints \eqref{real_time_mc_gen_0}--\eqref{real_time_mc_gen_2}, and the load shedding is bounded by constraints \eqref{real_time_load_shed_lims}.

The dual variables, stated after the colon signs in problems \eqref{prob:day_ahead_mc} and \eqref{prob:real_time_mc},  are used for pricing day-ahead contracts and penalizing/compensating power generation in real-time, respectively. \blue{We consider that the matrix of second-order cost coefficients $\mb{C}$ is positive-definite,  rendering} the objective functions \eqref{day_ahead_mc_obj} and \eqref{real_time_mc_obj} strictly \blue{convex} in primal decision variables. Hence, their primal and dual solutions are unique w.r.t. a particular wind power prediction $\widehat{\mb{w}}.$

\subsection{Revenue-Maximizing Wind Power Forecasting}\label{subsec:forecasting}
Consider a single wind power producer training a machine learning model $\mathbb{W}_{\mbg{\theta}}:\mathcal{F}\mapsto\mathcal{W}$ which maps features to power generation, i.e., $\mathbb{W}_{\mbg{\theta}}(\mbg{\varphi})=\widehat{w}$, where $\mbg{\theta}$ is a learning parameter, $\mbg{\varphi}$ is the vector of features, and $\widehat{w}$ is a prediction. The feature space $\mathcal{F}$ may include meteorological data (wind e.g., speed and direction), turbine data (e.g., the blade pitch angle), and their arbitrary transformations, such as polynomial feature transformations commonly used in the estimation of theoretical wind power curves \cite{wang2018wind}. Consider next a dataset $\mathcal{D}=\{(\mbg{\varphi}_{1},w_{1}),\dots,(\mbg{\varphi}_{n},w_{n})\}$ of $n$ historical samples. 
\begin{assumption}\label{ass:2}
    Samples size $n\gg|\mbg{\varphi}|.$ 
\end{assumption}
\blue{This standard assumption insures that we have more observations than features, allowing for model estimation by minimizing the following prediction loss function:}
\begin{align}
    \mathcal{L}(\mbg{\theta}\;\!|\;\!\mathcal{D})\vcentcolon=\frac{1}{n}\sum_{i=1}^{n}\norm{\mathbb{W}_{\mbg{\theta}}(\mbg{\varphi}_{i})-w_{i}}_{2}^{2}.\label{pred_loss_function}
\end{align}

Although the resulting model -- for a given model class and data -- may provide the minimum of prediction loss, it remains myopic to regulation penalties in the real-time market. In power systems with typically asymmetric costs, i.e., $|\mb{c}^{+}-\mb{c}|\ne|\mb{c}^{-}-\mb{c}|$ point-wise, wind power producers tune their forecasts to favor less expensive regulation and thus reduce real-time penalties \cite{bathurst2002trading}. The optimal forecast tuning can be formalized as a decision-focused learning problem.  Given the history of locational marginal prices at day-ahead and real-time markets, denoted by $\lambda_{1}$ and $\lambda_{2}$, respectively, a wind power producer optimizes model $\mathbb{W}_{\mbg{\theta}}$ by maximizing the following revenue function:
\begin{align}
    \mathcal{R}(\mbg{\theta}\;\!|\;\!\mathcal{D})\vcentcolon=&\;\frac{1}{n}\sum_{i=1}^{n}\Big(\lambda_{1i}\mathbb{W}_{\mbg{\theta}}(\mbg{\varphi}_{i}) +\lambda_{2i}(w_{i}-\mathbb{W}_{\mbg{\theta}}(\mbg{\varphi}_{i}))\Big)\label{prob:dfl_base}
\end{align}
where the first summand is the day-ahead revenue from submitted forecast, and the second summand is the real-time revenue.  \blue{Thanks to the cost order $\mb{c}^{+}>\mb{c}>\mb{c}^{-}$, the resulting locational marginal prices are such that over-prediction is always penalized in real-time with $\lambda_{2}>\lambda_{1}$, and under-prediction results in excess real-time power supply priced at $\lambda_{2}<\lambda_{1}$.} Hence, the wind power producer has financial incentives to sell all energy at the day-ahead market, thereby reducing real-time imbalances. However, the inherent erroneous nature of machine learning model $\mathbb{W}_{\mbg{\theta}}$ can not completely eliminate forecast errors and thus the real-time penalties. Hence, optimizing \eqref{prob:dfl_base}, a wind power producer finds the optimal learning parameter $\mbg{\theta}^{\star}$ that maximizes the revenue across the two markets on average. \blue{Selecting \eqref{prob:dfl_base} over \eqref{pred_loss_function}, the decision-maker implicitly controls the risk of high real-time penalties, which are not modeled in the standard loss function.}

Both prediction- and decision-focused objectives of wind power forecasting can be gathered in a single optimization:
\begin{align}\label{prob:mixed_obj}
    \maximize{\norm{\mbg{\theta}}_{1}\leqslant\tau}\quad\mathcal{R}(\mbg{\theta}\;\!|\;\!\mathcal{D}) - \gamma \mathcal{L}(\mbg{\theta}\;\!|\;\!\mathcal{D}),
\end{align}
which \blue{maximizes the revenue and minimizes the prediction loss weighted by some small parameter $\gamma > 0$. The presence of the loss term is justified as follows: if there are several regression models that bring the same level of revenues, pick the model that remains closer to the physical process underlying the data. This term also brings important regularizing properties to enable equilibrium analysis in Sec. \ref{subsec:ex_uniquness}.}


The variable $\mbg{\theta}$ of \eqref{prob:mixed_obj} is constrained to lie in a $\ell_{1}-$ball with $\tau>0$. This constraint forms a compact and non-empty set for the learning model --  the sufficient condition for the upcoming equilibrium existence result. It also allows for the optimal feature selection \cite{tibshirani1996regression} and solution robustness \cite{bertsimas2018characterization}, \blue{yet it must be chosen carefully to avoid affecting the value of the objective function.}

\section{Regression Equilibrium}

From an electricity market perspective, we are dealing with many wind power producers optimizing their regression models on market data. We thus seek an optimal {\it regression profile} $\mbg{\Theta}=(\mbg{\theta}_{1}^{\star},\dots,\mbg{\theta}_{b}^{\star})$ -- \blue{a set of $b$ private regression models} forming an equilibrium, such that no producer would benefit from a unilateral
deviation from their equilibrium forecast model. Such regression profile defines a pure Nash equilibrium, whose existence and computation we study in this section. We first formalize the optimization problems of market participants and provide equilibrium conditions in Sec. \ref{subsec:eqmodel}. We then study equilibrium properties in Sec. \ref{subsec:ex_uniquness} and its computation in Sec. \ref{subsec:compute}.  

\subsection{Regression Equilibrium Model}\label{subsec:eqmodel}
\begin{figure*}[b]

\hrule
\begin{subequations}\label{eq:revenues}
\begin{align}
 \mathcal{P}^{\text{W}}(\mbg{\Theta}\;\!|\;\!\mbg{\lambda}_{1},\mbg{\lambda}_{2},\mbg{\varphi},\mb{w})\vcentcolon=&\;\underbrace{\mbg{\lambda}_{1}^{\top}(\mbg{\Theta}\mbg{\varphi})}_{\text{day-ahead revenue}}\!\!\! + \;\underbrace{\mbg{\lambda}_{2}^{\top}(\mb{w}-\mbg{\Theta}\mbg{\varphi})}_{\text{real-time revenue}} -\underbrace{\norm{\mbg{\Theta}\mbg{\varphi}-\mb{w}}_{\mb{\Gamma}}^{2}}_{\text{regression loss}}\label{wind_producer_obj}\\
 \mathcal{P}^{\text{G}}(\mb{g}\;\!|\;\!\mbg{\lambda}_{1},\mbg{\lambda}_{2})\vcentcolon=&\;\underbrace{\mbg{\lambda}_{1}^{\top}\mb{p}}_{\text{day-ahead revenue}}\!\!\!+\;\;\;\underbrace{\mbg{\lambda}_{2}^{\top}(\mb{r}^{+}-\mb{r}^{-})}_{\text{real-time revenue}}-\underbrace{\left(\norm{\mb{p}+\mb{r}^{+}-\mb{r}^{-}}_{\mb{C}}^{2}+\mb{c}^{\top}\mb{p}+\mb{c}^{+\top}\mb{r}^{+}-\mb{c}^{-\top}\mb{r}^{-}\right)}_{\text{generation cost}}\label{conv_producer_obj}\\
  \mathcal{P}^{\text{L}}(\mbg{\ell}\;\!|\;\!\mbg{\lambda}_{2})\vcentcolon=&\;\underbrace{\mbg{\lambda}_{2}^{\top}\mbg{\ell}}_{\text{load shedding payment}}\!\!\!-\;\;\;
  \underbrace{\left(\norm{\mbg{\ell}}_{\mb{S}}+ \mb{s}^{\top}\mbg{\ell}\right)}_{\text{load shedding cost}}\label{el_load_obj}
\end{align}
\end{subequations}
\vspace{-0.15cm}
\hrule
\end{figure*}
Throughout the paper, we make the following assumptions. 
\begin{assumption}\label{ass:3}
    The class of private machine learning models $W_{\mbg{\theta}_{1}}(\mbg{\varphi}),\dots,W_{\mbg{\theta}_{b}}(\mbg{\varphi})$ is linear in feature vector $\mbg{\varphi}$. 
\end{assumption}
Before proceeding, we justify this assumption. First, although the models are linear in $\mbg{\varphi}$, the non-linearity of wind power generation can be captured by transformed features forming vector $\mbg{\varphi}$. Indeed, wind power generation is proportional to the cubed wind speed, which can be included in the feature vector $\mbg{\varphi}$, preserving the linearity of the model. More generally, the feature vector can be composed of non-linear kernel functions \cite{shawe2004kernel}, e.g., Gaussian kernels. Then, $\ell_{1}$-regularization, introduced in problem \eqref{prob:mixed_obj}, can identify the set of optimal kernels whose linear combination best explains the non-linear process underlying the data. Second, this assumption allows for co-optimizing the regression profile and market-clearing decisions in one convex optimization problem--the property we will leverage for computing equilibrium. Other convex models, such as input-convex neural networks \cite{amos2017input,dvorkin2023emission} are also of interest, yet to be addressed in future. \blue{Third, the combination of Assumptions \ref{ass:2} and \ref{ass:3} ensures the uniqueness of private regression models.}

\begin{assumption}\label{ass:4}
    Private models act on the same set of features.  
\end{assumption}

The assumption homogenizes private models, such that market outcomes will be explained by benefits of equilibrium and will not be obscured by heterogeneity of models or data. 

The equilibrium is formed by market participants, whose private optimization problems are explained next, followed by the details on equilibrium conditions. 

\subsubsection{Wind Power Producers} Each producer optimizes its own machine learning model for wind power forecasting. However, thanks to Assumptions \ref{ass:3} and \ref{ass:4} that homogenize forecast models of all producers, we can optimize the entire regression profile $\mbg{\Theta}$ using a single optimization. The goal of this optimization is to maximize the profit of wind power producers in response to electricity prices from day-ahead and real-time markets. For a particular wind power scenario, the profit function is provided in equation \eqref{wind_producer_obj}, which is similar to the objective function in \eqref{prob:mixed_obj} with exception that $\mathcal{P}^{\text{W}}$ is an aggregated profit. Here, matrix $\mbg{\Gamma}=\text{\blue{diag}}[\gamma_{1},\dots,\gamma_{b}]$ gathers the private weights of all producers. Then, the profit-maximizing problem for wind power producers is:
\begin{subequations}\label{prob:wind_producer}
\begin{align}
 \maximize{\mbg{\Theta}=(\mbg{\theta}_{1},\dots,\mbg{\theta}_{b})}\quad& \frac{1}{n}\sum_{i=1}^{n}\mathcal{P}^{\text{W}}(\mbg{\Theta}\;\!|\;\!\mbg{\lambda}_{1i},\mbg{\lambda}_{2i},\mbg{\varphi}_{i},\mb{w}_{i})
 \label{wind_producer_obj}\\\
 \st\quad&\mbg{\Theta}\in\mathcal{O}\vcentcolon=\mathcal{O}_{1}\times\dots\times\mathcal{O}_{b},%
\end{align}
\end{subequations}
that maximizes the average profit across the training datasets. The constraint set for private regression profiles is defined as
\begin{align}
    \mathcal{O}_{j}\vcentcolon=\left\{\mbg{\theta}\;\big|\norm{\mbg{\theta}}_{1}\leqslant\tau_{j},\right\},\quad\forall j=1,\dots,b.
\end{align}

For some feature vector $\mbg{\varphi}$, the profit-maximizing forecasts are obtained from the regression profile as $\widehat{\mb{w}}=\mbg{\Theta}^{\star}\mbg{\varphi}$.

\subsubsection{Conventional Generators} The aggregated profit function of conventional generators is provided in equation \eqref{conv_producer_obj}, which includes revenue streams from the day-ahead market for nominal power supply and from the real-time market for provided regulation. Here, vector $\mb{g}=(\mb{p},\mb{r}^{+},\mb{r}^{-})$ collects generator decisions in both markets. For each sample $i$ in the training dataset $\mathcal{D}$, $\mb{g}_{i}$ collects the corresponding dispatch generator decisions, and we use matrix $\mb{G}=(\mb{g}_{i},\dots,\mb{g}_{n})$ to collect decisions across the training dataset. The optimization problem for conventional producers then takes the form:
\begin{subequations}\label{prob:conv_producer}
\begin{align}
    \maximize{\mb{G}=(\mb{g}_{i},\dots,\mb{g}_{n})}\quad &
    \frac{1}{n}\sum_{i=1}^{n}\mathcal{P}^{\text{G}}(\mb{g}_{i}\;\!|\;\!\mbg{\lambda}_{1i},\mbg{\lambda}_{2i})\\
    \st\quad& \mb{G}\in\mathcal{G}
    \end{align}
\end{subequations}
where the constraint set $\mathcal{G}$ is defined as
\begin{align}
    \mathcal{G}\vcentcolon=\left\{
    \mb{G}\;
    \begin{array}{|l}
         \mb{\underline{p}}\leqslant\mb{p}_{i}+\mb{r}_{i}^{+}-\mb{r}_{i}^{-}\leqslant\mb{\overline{p}},\\
         \mb{0}\leqslant\mb{r}_{i}^{+}\leqslant\mb{\overline{p}}-\mb{p}_{i},\\
         \mb{0}\leqslant\mb{r}_{i}^{-}\leqslant\mb{p}_{i}-\mb{\underline{p}},\\
         \forall i=1,\dots,n
    \end{array}\right\}.
\end{align}

\blue{The modeling of conventional generators in \eqref{prob:conv_producer} reflects the realities of competitive electricity markets, where producers enter with their price-quantity bids and get cleared if the price covers their costs. In the competitive market setting, this simply means the best per-scenario response to day-ahead and real-time prices. In future work, we will consider conventional generators as ML users as well, who may optimize their ML models across market-clearing scenarios (e.g., to predict optimal self-commitment \cite{10530167}) in a similar spirit as wind power producers do in problem \eqref{prob:wind_producer}.} 

\subsubsection{Loads} Whenever wind and conventional generators lack capacity to satisfy electricity demands, the loads can shed a part of their consumption at a very high cost to the system. The aggregated profit function for loads in this case is given in equation \eqref{el_load_obj}, and their 
profit-maximizing problem is 
\begin{subequations}
\begin{align}
    \maximize{\mb{L}=(\mbg{\ell}_{i},\dots,\mbg{\ell}_{n})}\quad &
    \frac{1}{n}\sum_{i=1}^{n}\mathcal{P}^{\text{L}}(\mbg{\ell}_{i}\;\!|\;\!\mbg{\lambda}_{2i})\\
    \st\quad& \mb{L}\in\mathcal{S}
    \end{align}
\end{subequations}
where $\mathcal{S}$ is the constraint set for load shedding decisions, i.e.,
\begin{align}
    \mathcal{S}\vcentcolon=\left\{
    \mb{L}\;
    \begin{array}{|l}
         \mb{0}\leqslant\mbg{\ell}_{i}\leqslant\mb{d},\\
         \forall i=1,\dots,n
    \end{array}\right\}.
\end{align}

\subsubsection{Equilibrium Conditions} These conditions build upon market settings in Sec. \ref{subsec:setting}. They couple decision-making of market participants and yield equilibrium-supporting prices. They first require power balance at day-ahead and real-time markets via the following complementarity conditions: 
\begin{subequations}\label{eq_cond}
\begin{align}
    &0 \leqslant \mu_{1i} \perp \mb{1}^{\top}(\mb{p}_{i} +\mbg{\Theta} \mbg{\varphi}_{i} - \mb{d})  \geqslant 0, \label{eq_cond_1}\\
    &0 \leqslant  \mu_{2i} \perp \mb{1}^{\top}(\mb{r}_{i}^{+}-\mb{r}_{i}^{-} - \mbg{\Theta} \mbg{\varphi}_{i}+\mb{w}_{i}+\mbg{\ell}_{i}) \geqslant 0,
\end{align}
where $\mu_{1i}$ and $\mu_{2i}$ are prices used to support power balance. \blue{If the power balance in either of the two markets is satisfied with equality (e.g., as in (1b) and (2b)), then there exists a corresponding non-negative price that compensates generation for maintaining sufficient output, which keeps the system in balance. If the power balance does not hold as an equality, the corresponding price will be zero (market failure). The remaining conditions are complementarities associated with the satisfaction of network limits in the day-ahead market} The remaining conditions are complementarities associated with the satisfaction of network limits in the day-ahead market
\begin{align}
    &\mb{0} \leqslant  \overline{\mbg{\kappa}}_{1i} \perp \mb{\overline{f}} -  \mb{F}(\mb{p}_{i} + \mbg{\Theta} \mbg{\varphi}_{i} - \mb{d}) \geqslant \mb{0}, \\
    &\mb{0} \leqslant  \underline{\mbg{\kappa}}_{1i} \perp   \mb{\overline{f}} +  \mb{F}(\mb{p}_{i} + \mbg{\Theta} \mbg{\varphi}_{i} - \mb{d}) \geqslant \mb{0},
\end{align}
and in the real-time market:
\begin{align}
   \mb{0}\leqslant \overline{\mbg{\kappa}}_{2i} \perp \mb{\overline{f}} \;\!-\;\!&\mb{F}(\mb{p}_{i} + \mbg{\Theta} \mbg{\varphi}_{i} - \mb{d}) \nonumber\\ 
   \;\!-\;\!&\mb{F}(\mb{r}_{i}^{+}-\mb{r}_{i}^{-} - \mbg{\Theta} \mbg{\varphi}_{i}+\mb{w}_{i}+\mbg{\ell}_{i}) \geqslant \mb{0},\\
   \mb{0}\leqslant  \underline{\mbg{\kappa}}_{2i} \perp \mb{\overline{f}} \;\!+\;\!&\mb{F}(\mb{p}_{i} + \mbg{\Theta} \mbg{\varphi}_{i} - \mb{d}) \nonumber\\ 
   \;\!+\;\!&\mb{F}(\mb{r}_{i}^{+}-\mb{r}_{i}^{-} - \mbg{\Theta} \mbg{\varphi}_{i}+\mb{w}_{i}+\mbg{\ell}_{i})\geqslant \mb{0},
\end{align}
\end{subequations}
which similarly state that if the power flow constraints are satisfied with equality, as in case of congestion, there exists non-negative equilibrium prices in the corresponding market.  

Dual variables $\mu_{1i},\mu_{2i},\overline{\mbg{\kappa}}_{1i},\underline{\mbg{\kappa}}_{1i},\overline{\mbg{\kappa}}_{2i},\underline{\mbg{\kappa}}_{2i},\forall i=1,\dots,n,$ form location marginal prices. \blue{Drawing from the differentiating of the partial Lagrangian function of the market-clearing optimization problem in} \cite[\blue{Problem (1)--(4)}]{litvinov2010design}, these prices are derived by relating dual variables to the day-ahead and real-time energy volumes, resulting in
\begin{subequations}\label{LMPs}
\begin{align}
    \mbg{\lambda}_{1i} \vcentcolon=&\; \mu_{1i}\cdot\mb{1} - \mb{F}^{\top}\!\left(\overline{\mbg{\kappa}}_{1i}-\underline{\mbg{\kappa}}_{1i}+\overline{\mbg{\kappa}}_{2i}-\underline{\mbg{\kappa}}_{2i}\right),\\
    \mbg{\lambda}_{2i} \vcentcolon=&\; \mu_{2i}\cdot\mb{1} - \mb{F}^{\top}\!\left(\overline{\mbg{\kappa}}_{2i}-\underline{\mbg{\kappa}}_{2i}\right).
\end{align}
\end{subequations}
They are used in the profit functions of market participants in equations \eqref{wind_producer_obj} through \eqref{el_load_obj}, respectively. Importantly, as these prices come from the complementarity conditions above, they must be seen as functions of market participant decisions, and the regression profile $\mbg{\Theta}=(\mbg{\theta}_{1},\dots,\mbg{\theta}_{b})$ in particular. 

\subsubsection{The Model} The Nash Equilibrium (NE) model consists of the following profit-maximizing problems:
\begin{align}
&\left\{
\begin{aligned}
    &\maximize{\mbg{\Theta}\in\mathcal{O}}\quad \frac{1}{n}\sum_{i=1}^{n}\mathcal{P}^{\text{W}}(\mbg{\Theta}\;\!|\;\!\mbg{\lambda}_{1i},\mbg{\lambda}_{2i},\mbg{\varphi}_{i},\mb{w}_{i})\\
    &\maximize{\mb{G}\in\mathcal{G}}\quad 
    \frac{1}{n}\sum_{i=1}^{n}\mathcal{P}^{\text{G}}(\mb{g}_{i}\;\!|\;\!\mbg{\lambda}_{1i},\mbg{\lambda}_{2i})\\
    &\maximize{\mb{L}\in\mathcal{S}}\quad
    \frac{1}{n}\sum_{i=1}^{n}\mathcal{P}^{\text{L}}(\mbg{\ell}_{i}\;\!|\;\!\mbg{\lambda}_{2i})
\end{aligned}
\right\}\label{NE}
\end{align}
where the prices $\mbg{\lambda}_{11},\dots,\mbg{\lambda}_{1n}$ and $\mbg{\lambda}_{21},\dots,\mbg{\lambda}_{2n}$ are chosen such that the following complementarities are satisfied:
\begin{subequations}\label{comp_cond}
\begin{align}
    0\leqslant\mu_{1i}\perp&\; \mb{1}^{\top}(\mb{p}_{i} +\mbg{\Theta} \mbg{\varphi}_{i} - \mb{d}) \geqslant0\\
    0\leqslant\mu_{2i}\perp&\; \mb{1}^{\top}(\mb{r}_{i}^{+}-\mb{r}_{i}^{-} - \mbg{\Theta} \mbg{\varphi}_{i}+\mb{w}_{i}+\mbg{\ell}_{i}) \geqslant 0\\
    \mb{0}\leqslant\overline{\mbg{\kappa}}_{1i}\perp&\; \mb{\overline{f}} -  \mb{F}(\mb{p}_{i} + \mbg{\Theta} \mbg{\varphi}_{i} - \mb{d}) \geqslant \mb{0}\\
    \mb{0}\leqslant\underline{\mbg{\kappa}}_{1i}\perp&\; \mb{\overline{f}} +  \mb{F}(\mb{p}_{i} + \mbg{\Theta} \mbg{\varphi}_{i} - \mb{d}) \geqslant \mb{0}\\
    \mb{0}\leqslant\overline{\mbg{\kappa}}_{2i}\perp&\; \mb{\overline{f}} -\mb{F}(\mb{p}_{i} + \mbg{\Theta} \mbg{\varphi}_{i} - \mb{d})\nonumber\\
    &\quad\quad\!-\mb{F}(\mb{r}_{i}^{+}-\mb{r}_{i}^{-} - \mbg{\Theta} \mbg{\varphi}_{i}+\mb{w}_{i}+\mbg{\ell}_{i}) \geqslant \mb{0}\\
    \mb{0}\leqslant\underline{\mbg{\kappa}}_{2i}\perp&\; \mb{\overline{f}} +\mb{F}(\mb{p}_{i} + \mbg{\Theta} \mbg{\varphi}_{i} - \mb{d})\nonumber\\
    &\underbrace{\quad\quad\!+\mb{F}(\mb{r}_{i}^{+}-\mb{r}_{i}^{-} - \mbg{\Theta} \mbg{\varphi}_{i}+\mb{w}_{i}+\mbg{\ell}_{i}) \geqslant \mb{0}}_{\text{shared constraint set} \;\mathcal{E}}
\end{align}
\end{subequations}
for every sample $i=1,\dots,n$ of the training dataset $\mathcal{D}$.

\subsection{Existence and Uniqueness of Regression Equilibrium}\label{subsec:ex_uniquness}

Problem \eqref{NE}-\eqref{comp_cond} is an instance of NE problem with shared constraints. To establish the existence and uniqueness of the equilibrium, we recast it as a {\it variational inequality} (VI) problem \cite{facchinei2003finite}, and then rest on the well-developed theory of VIs. The VI problem is defined as follows.
\begin{definition} Given a closed and convex set $\mathcal{K}\subseteq\mathbb{R}^{k}$ and a mapping $F:\mathcal{K}\mapsto\mathbb{R}^{k}$, the VI problem, denoted $\text{VI}(\mathcal{K},F)$, consists in finding an optimal profile $(\mbg{\Theta}^{\star},\mb{G}^{\star}, \mb{L}^{\star})\in\mathcal{K}$, called a solution to the VI problem, such that $$
\left(\begin{bmatrix*}[r]
\mbg{\Theta}\\\mb{G}\\\mb{L}
\end{bmatrix*}
-
\begin{bmatrix*}[r]
\mbg{\Theta}^{\star}\\\mb{G}^{\star}\\\mb{L}^{\star}
\end{bmatrix*}\right)
^{\top}F\left(
\mbg{\Theta}^{\star},\mb{G}^{\star},\mb{L}^{\star}
\right)\geqslant0,$$
for all $(\mbg{\Theta},\mb{G},\mb{L})\in\mathcal{K}$.
\end{definition}
The NE problem \eqref{NE}--\eqref{comp_cond} can recast as VI when the constraint set $\mathcal{K}$ is an intersection of feasible regions of market participants and the shared constraints $\mathcal{E}$. 
\begin{assumption}\label{ass:5}
    Set $\mathcal{K}\vcentcolon=\mathcal{O}\cap\mathcal{G}\cap\mathcal{S}\cap\mathcal{E}$ is compact.
\end{assumption}
This assumption simply requires that electricity demands $\mb{d}$, power transfer distribution factors $\mb{F}$, transmission capacities $\overline{\mb{f}}$, and actual wind power generation $\mb{w}_{1},\dots,\mb{w}_{n}$ are such that  there exists at least one regression, generation and load shedding profile satisfying private and equilibrium constraints simultaneously. \blue{Following the discssion on page 46 in \cite{scutari2010convex}, the underlying mapping $F$ formulates by stacking the partial derivatives of the private objective functions excluding the terms related to the shared constraints, i.e.,}
\begin{align}
    F(\mbg{\Theta},\mb{G},\mb{L}) = 
    \begin{bmatrix*}[l]
        \\[-0.8em]
        \nabla_{\mbg{\Theta}}\frac{1}{n}\textstyle\sum_{i=1}^{n}\mathcal{P}^{\text{W}}(\mbg{\Theta}\;\!|\;\!\mb{0},\mb{0},\mbg{\varphi}_{i},\mb{w}_{i})\\[1em]
        \nabla_{\mb{g}_{1}}\frac{1}{n}\;\!\mathcal{P}^{\text{G}}\!\;\!\;(\mb{g}_{1}\;\!|\;\!\mb{0},\mb{0})\\[-0.25em]
        \dots\\[-0.15em]
        \nabla_{\mb{g}_{n}}\frac{1}{n}\mathcal{P}^{\text{G}}\!\;\!\;(\mb{g}_{n}\;\!|\;\!\mb{0},\mb{0})\\[1em]
        \nabla_{\mbg{\ell}_{1}}\frac{1}{n}\;\!\mathcal{P}^{\text{L}}\!\;\!\;(\mbg{\ell}_{1}\;\!|\;\!\mb{0})\\[-0.25em]
        \dots\\[-0.15em]
        \nabla_{\mbg{\ell}_{n}}\frac{1}{n}\mathcal{P}^{\text{L}}\!\;\!\;(\mbg{\ell}_{n}\;\!|\;\!\mb{0})\\
    \end{bmatrix*},\label{grad_map}
\end{align}
\blue{The equilibrium prices in \eqref{grad_map} are set to zeros because they are not variables of the VI problem. However, since the shared constraints $\mathcal{E}$ participate in set $\mathcal{K}$, the VI solution will satisfy $\mathcal{E}$. Such correspondence between \eqref{NE}-\eqref{comp_cond} and $\text{VI}(\mathcal{K},F)$  means that if $(\mbg{\Theta}^{\star},\mb{G}^{\star},\mb{L}^{\star})$ solves $\text{VI}(\mathcal{K},F),$ then there exists a set of prices  $\mbg{\lambda}^{\star}=(\mbg{\lambda}_{11}^{\star},\dots,\mbg{\lambda}_{1n}^{\star},\mbg{\lambda}_{21}^{\star},\dots,\mbg{\lambda}_{2n}^{\star})$, such that $(\mbg{\Theta}^{\star},\mb{G}^{\star},\mb{L}^{\star},\mbg{\lambda}^{\star})$ is the solution to the NE problem \eqref{NE}-\eqref{comp_cond}; conversely, if $(\mbg{\Theta}^{\star},\mb{G}^{\star},\mb{L}^{\star},\mbg{\lambda}^{\star})$ is the solution to \eqref{NE}-\eqref{comp_cond}, then $(\mbg{\Theta}^{\star},\mb{G}^{\star},\mb{L}^{\star})$ is the solution to $\text{VI}(\mathcal{K},F)$. We refer to \cite{scutari2010convex} for more examples of NE problems with shared constraints and their reformulation as VI problems.}

The correspondence between \eqref{NE}-\eqref{comp_cond} and $\text{VI}(\mathcal{K},F)$ allows us to establish the existence of the NE from the existence of a solution to the VI. Furthermore, the uniqueness of the VI solution also implies the uniqueness of the NE, as shown via the following result.

\begin{theorem}\label{th:ne_existence_uniqness}
    For the NE problem \eqref{NE}--\eqref{comp_cond} resting on Assumptions \ref{ass:1}--\ref{ass:4}, suppose that Assumption \ref{ass:5} also holds. Then there exists a unique equilibrium profile $(\mbg{\Theta}^{\star},\mb{G}^{\star},\mb{S}^{\star})$, such that for every wind power producer $j=1,\dots,b$, we have: 
    \begin{align}\label{eq_inequality}
        &\mathcal{P}^{\text{W}}(\mbg{\theta}_{j}\;\!|\;\!\mbg{\lambda}(\mbg{\theta}_{j},\mbg{\Theta}_{\!-\!j}^{\star},\mb{G}^{\star},\mb{S}^{\star}),\mathcal{D})\nonumber\\
        &\quad\quad\quad\leqslant\mathcal{P}^{\text{W}}(\mbg{\theta}_{j}^{\star}\;\!|\;\!\mbg{\lambda}(\mbg{\theta}_{j}^{\star},\mbg{\Theta}_{\!-\!j}^{\star},\mb{G}^{\star},\mb{S}^{\star}),\mathcal{D}),\forall \mbg{\theta}_{j}\in\mathcal{O}_{j}
    \end{align}
\end{theorem}
The proof of this result is relegated to Appendix \ref{app:one}.  One major implication is that there exists a unique regression equilibrium profile $\boldsymbol{\Theta}^{\star}$ among wind power producers, such that any unilateral deviation from the equilibrium regression would not results in extra profits, in a competitive setting. 

\begin{remark}
    The result of Theorem \ref{th:ne_existence_uniqness} holds on the training dataset. In Section \ref{sec:experiments}, we will demonstrate how the equilibrium regression profile generalizes on the testing dataset as well.
\end{remark}

\subsection{Computing Regression Equilibrium}\label{subsec:compute}

An important upshot of Theorem \ref{th:ne_existence_uniqness} is the existence of a centralized optimization problem which solves the equilibrium.

\begin{corollary}\label{corollary}
    Under Assumptions \ref{ass:1} to \ref{ass:5}, the following regularized cost-minimization problem solves the NE \eqref{NE}--\eqref{comp_cond}:
\vspace{-1em}
\begin{align*}
\minimize{\mbg{\Theta},\mb{G},\mb{L}}\quad&\frac{1}{n}\sum_{i=1}^{n}\Big(\norm{\mb{p}_{i}+\mb{r}_{i}^{+}-\mb{r}_{i}^{-}}_{\mb{C}}^{2}+\mb{c}^{\top}\mb{p}_{i}+\mb{c}^{+\top}\mb{r}_{i}^{+} \nonumber \\
&-\mb{c}^{-\top}\mb{r}_{i}^{-} + \norm{\mbg{\ell}_{i}}_\mb{S}^{2}+\mb{s}^{\top}\mbg{\ell}_{i}+ \norm{\mbg{\Theta} \mbg{\varphi}_{i}-\mb{w}_{i}}_{\mbg{\Gamma}}^{2}\Big)\nonumber\\
\st\quad&\mb{1}^{\top}(\mb{p}_{i} +\mbg{\Theta} \mbg{\varphi}_{i} - \mb{d}) = 0, \\
&\mb{1}^{\top}(\mb{r}_{i}^{+}-\mb{r}_{i}^{-} - \mbg{\Theta} \mbg{\varphi}_{i}+\mb{w}_{i}+\mbg{\ell}_{i}) = 0, \\
&|\mb{F}(\mb{p}_{i} + \mbg{\Theta} \mbg{\varphi}_{i} - \mb{d})| \leqslant \mb{\overline{f}},\\
&|\mb{F}(\mb{p}_{i} + \mbg{\Theta} \mbg{\varphi}_{i} - \mb{d}) \nonumber\\
&\quad+\mb{F}(\mb{r}_{i}^{+}-\mb{r}_{i}^{-} - \mbg{\Theta} \mbg{\varphi}_{i}+\mb{w}_{i}+\mbg{\ell}_{i})| \leqslant \mb{\overline{f}},\\
&\mb{\underline{p}}\leqslant\mb{p}_{i}+\mb{r}_{i}^{+}-\mb{r}_{i}^{-}\leqslant\mb{\overline{p}},\\
&\mb{0}\leqslant\mb{r}_{i}^{+}\leqslant\mb{\overline{p}}-\mb{p}_{i},\\
&\mb{0}\leqslant\mb{r}_{i}^{-}\leqslant\mb{p}_{i}-\mb{\underline{p}},\\
&\mb{0}\leqslant\mbg{\ell}_{i}\leqslant\mb{d},\quad\quad\quad\;\;\;\;\! \forall i=1,\dots,n,\\
&\norm{\mbg{\theta}_{j}}_{1}\leqslant\tau_{j},\;\quad\quad\quad\;\;\!\forall j=1,\dots,k. 
\end{align*}
\end{corollary}
The objective here is to minimize the average social cost of electricity, regularized by the total regression loss up to chosen parameter $\mbg{\Gamma}.$ The feasible region includes the day-ahead and real-time market-clearing constraints, enforced on each sample $i$ of the training dataset $\mathcal{D}$, as well as $\ell_{1}-$regularization constraints on regression parameters. With a small $\mbg{\Gamma}$, the optimal regression profile $\mbg{\Theta}^{\star}$ minimizes the cost across the day-ahead and real-time markets on average, thus improving the coordination between the day-ahead and real-time markets. 

\blue{We owe the existence of centralized optimization to the risk neutrality of market participants. It is possible to incorporate risk measures to model risk aversion, yet the existence of centralized optimization is not guaranteed without additional financial instruments. We refer to \cite{gerard2018risk} for details.}  

Alternatively to centralized optimization, we can also compute the equilibrium using an ADMM-based algorithm, which consists of several steps:

\textit{\bf Step 0:} Choose any $\mbg{\lambda}_{1i}^{0}\geqslant0$ and $\mbg{\lambda}_{2i}^{0}\geqslant0$ for each sample $i=1,\dots,n$ of the training dataset, select a step size $\varrho>0$. and set iteration index $k\leftarrow0$.

\textit{\bf Step 1:} If LMPs $\mbg{\lambda}_{11}^{k},\dots,\mbg{\lambda}_{1n}^{k}$ and $\mbg{\lambda}_{21}^{k},\dots,\mbg{\lambda}_{2n}^{k}$ satisfy a suitable termination criteria, then stop. 

\textit{\bf Step 2:} Given LMPs $\mbg{\lambda}_{11}^{k},\dots,\mbg{\lambda}_{1n}^{k}$ and $\mbg{\lambda}_{21}^{k},\dots,\mbg{\lambda}_{2n}^{k}$, solve the augmented NE $(\mbg{\Theta}^{k},\mb{G}^{k},\mb{L}^{k})$ with fixed prices:
\vspace{-5pt}
\begin{subequations}\label{best_response}
\begin{align}
    &\mbg{\Theta}^{k}=\myargmax{\mbg{\Theta}\in\mathcal{O}}\; \frac{1}{n}\sum_{i=1}^{n}\overline{\mathcal{P}}_{\varrho}^{\text{W}}(\mbg{\Theta}\;\!|\;\!\mbg{\lambda}_{1i}^{k},\mbg{\lambda}_{2i}^{k},\mbg{\varphi}_{i},\mb{w}_{i},\mb{g}_{i}^{k-1}\!,\mbg{\ell}_{i}^{k-1})\label{best_response_wind}\\[-10.5pt]
    &\mb{G}^{k}=\myargmax{\mb{G}\in\mathcal{G}}\; 
    \frac{1}{n}\sum_{i=1}^{n}\overline{\mathcal{P}}_{\varrho}^{\text{G}}(\mb{g}_{i}\;\!|\;\!\mbg{\lambda}_{1i}^{k},\mbg{\lambda}_{2i}^{k},\mbg{\Theta}^{k-1},\mbg{\ell}_{i}^{k-1})\\
    &\mb{L}^{k}=\myargmax{\mb{L}\in\mathcal{S}}\;
    \frac{1}{n}\sum_{i=1}^{n}\overline{\mathcal{P}}_{\varrho}^{\text{L}}(\mbg{\ell}_{i}\;\!|\;\!\mbg{\lambda}_{2i}^{k},,\mbg{\Theta}^{k-1},\mb{g}_{i}^{k-1})
\end{align}
\end{subequations}

\textit{\bf Step 3:} Update the dual variables of complementarity conditions \eqref{comp_cond}  for all $i=1,\dots,n$ samples in the dataset:
\vspace{-14pt}
\begin{subequations}\label{dual_price}
\begin{align}
    \mu_{1i}^{k+1} \!\leftarrow&\! \left[\mu_{1i}^{k} - \varrho \cdot \mb{1}^{\top}(\mb{p}_{i}^{k} + \mbg{\Theta}^{k} \mbg{\varphi}_{i} - \mb{d})\right]_{+}\\
    \overline{\mbg{\kappa}}_{1i}^{k+1} \!\leftarrow&\!\left[\overline{\mbg{\kappa}}_{1i}^{k}-\varrho\cdot\left(\mb{\overline{f}} -  \mb{F}(\mb{p}_{i}^{k} + \mbg{\Theta}^{k} \mbg{\varphi}_{i} - \mb{d})\right)\right]_{+}\\
    \underline{\mbg{\kappa}}_{1i}^{k+1} \!\leftarrow&\!\left[\underline{\mbg{\kappa}}_{1i}^{k}-\varrho\cdot\left(\mb{\overline{f}} +  \mb{F}(\mb{p}_{i}^{k} + \mbg{\Theta}^{k} \mbg{\varphi}_{i} - \mb{d})\right)\right]_{+}\\
    \mu_{2i}^{k+1} \!\leftarrow&\! \left[\mu_{2i}^{k}\!-\! \varrho \cdot \mb{1}^{\top}(\mb{r}_{i}^{+,k}\!\!-\mb{r}_{i}^{-,k}\!\!- \mbg{\Theta}^{k} \mbg{\varphi}_{i}+\mb{w}_{i}+\mbg{\ell}_{i}^{k})\right]_{+}\hspace{-1em}\\
    \overline{\mbg{\kappa}}_{2i}^{k+1} \!\leftarrow&\Big[\overline{\mbg{\kappa}}_{2i}^{k}-\varrho\cdot\Big(\mb{\overline{f}} -  \mb{F}(\mb{p}_{i}^{k} + \mbg{\Theta}^{k} \mbg{\varphi}_{i} - \mb{d})\nonumber\\
    &\quad\quad\;\;\!\;\!\;\;\!\!\!-\mb{F}(\mb{r}_{i}^{+,k}\!\!-\mb{r}_{i}^{-,k}\!\!- \mbg{\Theta}^{k} \mbg{\varphi}_{i}+\mb{w}_{i}+\mbg{\ell}_{i}^{k})\Big)\Big]_{+}\hspace{-1em}\\
    \underline{\mbg{\kappa}}_{2i}^{k+1} \!\leftarrow&\Big[\underline{\mbg{\kappa}}_{2i}^{k}-\varrho\cdot\Big(\mb{\overline{f}} +  \mb{F}(\mb{p}_{i}^{k} + \mbg{\Theta}^{k} \mbg{\varphi}_{i} - \mb{d})\nonumber\\
    &\quad\quad\;\;\!\;\!\;\;\!\!\!+\mb{F}(\mb{r}_{i}^{+,k}\!\!-\mb{r}_{i}^{-,k}\!\!- \mbg{\Theta}^{k} \mbg{\varphi}_{i}+\mb{w}_{i}+\mbg{\ell}_{i}^{k})\Big)\Big]_{+}\hspace{-1em}
\end{align}
\end{subequations}
and then update electricity LMPs for all $i=1,\dots,n:$
\begin{subequations}\label{LMP_update}
\begin{align}
    \mbg{\lambda}_{1i}^{k+1}\!\leftarrow&\mu_{1i}^{k+1}\!\cdot\!\mb{1} - \mb{F}^{\top}\!\left(\overline{\mbg{\kappa}}_{1i}^{k+1}\!-\!\underline{\mbg{\kappa}}_{1i}^{k+1}\!+\!\overline{\mbg{\kappa}}_{2i}^{k+1}\!-\!\underline{\mbg{\kappa}}_{2i}^{k+1}\right),\\
    \mbg{\lambda}_{2i}^{k+1}\!\leftarrow&\mu_{2i}^{k+1}\!\cdot\!\mb{1} - \mb{F}^{\top}\!\left(\overline{\mbg{\kappa}}_{2i}^{k+1}\!-\!\underline{\mbg{\kappa}}_{2i}^{k+1}\right).
\end{align}
\end{subequations}

\textit{\bf Step 4:} Set $k\leftarrow k+1$ and go to Step 1. \\ 

At every iteration $k$, the algorithm computes the best response of market participants to electricity prices in the day-ahead and real-time markets by solving the augmented NE in \eqref{best_response}. This problem is similar to the NE in \eqref{NE} with a difference that the private profit-maximizing optimizations are augmented with the ADMM terms related to the feasibility of shared constraints in \eqref{comp_cond}. For example, the augmented profit function of wind power producers takes the form:
\begin{align}
    \overline{\mathcal{P}}_{\varrho}^{\text{W}}&(\mbg{\Theta}\;\!|\;\!\mbg{\lambda}_{1i},\mbg{\lambda}_{2i},\mbg{\varphi}_{i},\mb{w}_{i},\mb{g}_{i}^{k-1},\mbg{\ell}_{i}^{k-1}) \vcentcolon=\nonumber\\
    & \mathcal{P}^{\text{W}}(\mbg{\Theta}\;\!|\;\!\mbg{\lambda}_{1i},\mbg{\lambda}_{2i},\mbg{\varphi}_{i},\mb{w}_{i})\nonumber\\
    &\quad+\frac{\varrho}{2}\norm{\mb{1}^{\top}(\mb{p}_{i}^{k-1} + \mbg{\Theta} \mbg{\varphi}_{i} - \mb{d})}_{2}^{2}\nonumber\\
    &\quad\quad+\frac{\varrho}{2}\norm{\text{max}\left\{\mb{F}(\mb{p}_{i}^{k-1} + \mbg{\Theta} \mbg{\varphi}_{i} - \mb{d}) - \overline{\mb{f}},\mb{0}\right\}}_{2}^{2}\nonumber\\
    &\quad\quad\quad+\dots\footnotemark\label{admm_wind_problem}
\end{align}
\footnotetext{In the interest of space, formulation \eqref{admm_wind_problem} omits ADMM terms related to the remaining constraints in \eqref{comp_cond}, which are constructed by analogy.}where the augmented terms regularize the regression profile $\mbg{\Theta}$ to ensure the feasibility of day-ahead and real-time market constraints. Then, the best response is used for updating the dual variables via \eqref{dual_price}, where the current value is adjusted to account for either power balance or flow constraint violations using a small step size $\varrho$. These updates include a projection onto the non-negative orthant to satisfy the complementarity conditions \eqref{comp_cond}. These variables are subsequently used for LMP updates \eqref{LMP_update}, then conveyed to the next iteration. The algorithm terminates when it meets a prescribed convergence criterion for LMPs, e.g., verifying whether their changes do not exceed some tolerance on average or for each sample in the training data. The convergence of ADMM algorithms is well-established, as discussed in \cite{boyd2011distributed}.

\section{Numerical Experiments}\label{sec:experiments}
\subsection{Data and Settings}
The wind power data is taken from \cite{Dataset} and includes $2\sfrac{1}{2}$ years of readings. The readings are taken every 10 minutes and include the active power output, \blue{which is normalized in the range from 0 and 1}, and corresponding weather features, among which the wind speed and wind direction are selected for forecasting. \blue{The wind features are normalized using the min-max normalization, i.e., $x_{\text{norm}}(x)=\frac{x-x_{\text{min}}}{x_{\text{max}}-x_{\text{min}}}$.} We sampled $5,000$ and $10,000$ readings for training and testing, respectively, at random. We train a linear kernel regression with a Gaussian kernel $\varphi(x)=\text{exp}(-\varsigma\norm{\mu-x}_{\blue{2}}^{\blue{2}})$, \blue{where $\varsigma$ is the scale and $\mu$ is the center point. Each feature is transformed using 15 such kernels with their center points $\mu$ being equally spaced out in the feature domain. We empirically verified that 15 kernels per feature suffice to capture the non-linear characteristics of the wind power curve without overfitting the model.} As a result, for each wind power record, we have $|\boldsymbol{\varphi}|=30$ transformed features. To select only important feature transformations, $\ell_{1}$ regularization with $\tau=10$ is used. We use the IEEE $24-$Bus Reliability Test System \blue{from \cite{ordoudis2016updated} with six wind farms of $200$ MW capacity each, at buses $3,5,7,16,21$ and $23$, covering $45.3\%$} of the electricity demand when producing at maximum capacity. All wind farms are granted the same dataset and features so that the private regression models are homogeneous.
Generation cost are such that \blue{$\mb{c}^{+}=3\mb{c}$ and $\mb{c}^{-}=0.5\mb{c}$}, and the remaining network parameters ensure that there is no need to shed loads, so we solely discuss the normal operation. In our experiments, we compare three regression models:
\begin{itemize}
    \item {\bf Oracle:} Provides perfect predictions for benchmarking forecasts and discussing the benefits of equilibrium.
    \item {\bf Baseline:} A standard approach to forecasting, when wind power producers minimize the prediction loss \eqref{pred_loss_function}. 
    \item {\bf Equilibrium:} Provides solution to the NE problem \eqref{NE}--\eqref{comp_cond}, with revenue-seeking wind power producers optimizing their forecast models from market data. The weights balancing the prediction loss and revenues are set uniformly to $\gamma=10^{-4}$ for all wind power producers. 
\end{itemize}

In the next section, we first present the economic benefits of equilibrium, and then discuss computational aspects. 

\subsection{Results}
\begin{figure*}
    \centering
    \includegraphics[width=1\textwidth]{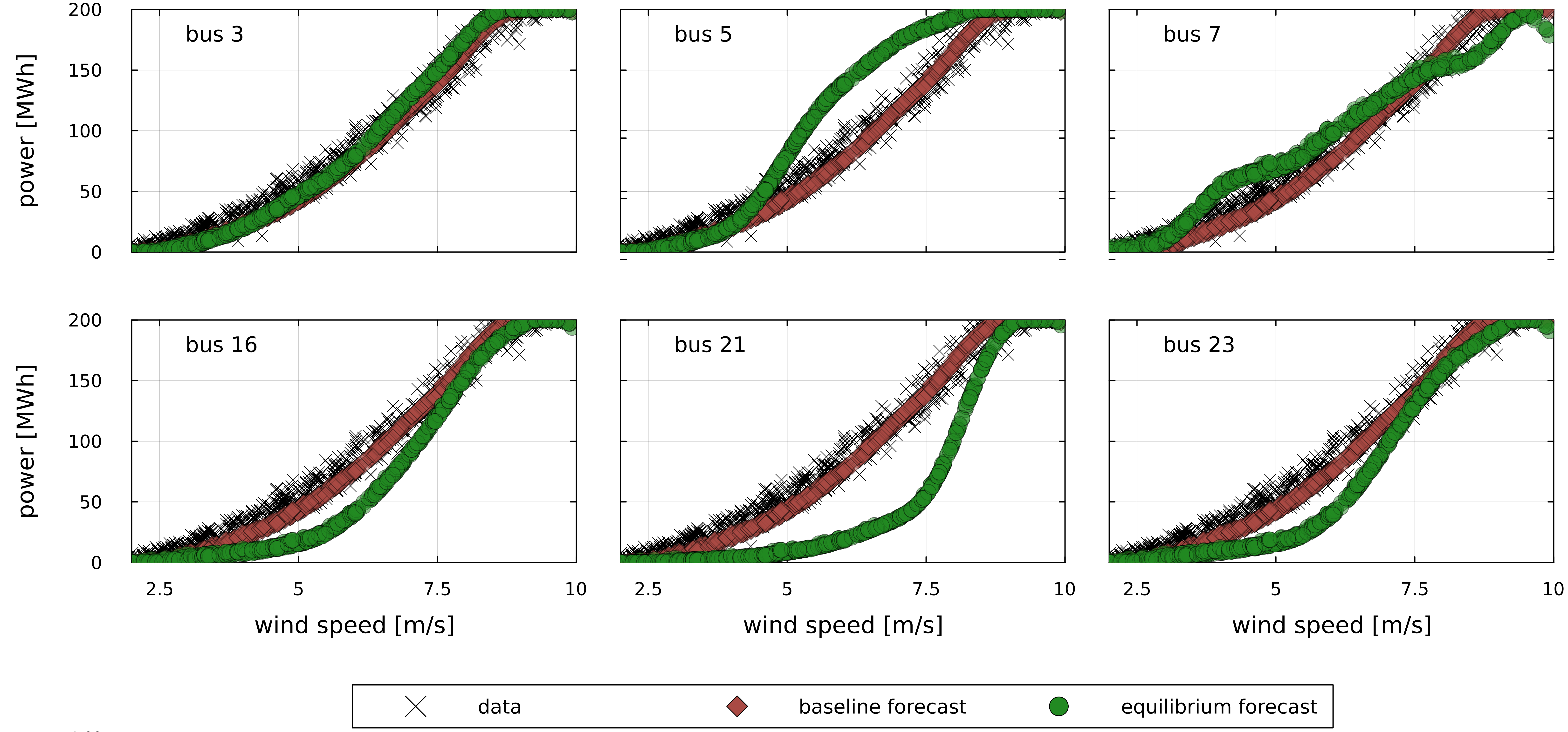}
    \caption{\blue{Baseline and equilibrium forecasts as functions of wind speed at six wind farms installed in the IEEE 24-Bus RTS.}}
    \label{fig:prediction}
\end{figure*}

The wind power forecasts resulting from the baseline and equilibrium regression models are depicted in Fig. \ref{fig:prediction}. Since the baseline regression solely optimizes for prediction loss, the resulting forecasts adequately explain the trends in data and are identical for all six wind farms. The equilibrium forecasts, on the other hand, optimize for private welfare and may notably deviate from the underlying data for each producer differently, as further shown in Fig. \ref{fig:prediction}. Such deviations from the ground truth data are driven by optimization objectives and are well-known in the literature. For example, the results of \cite{zhang2023deriving} and \cite{zhang2023value} show that the cost-optimal forecast consistently under-predicts the wind power output in the day-ahead market. Here, however, we account for network congestion, revealing that system-optimal under- or over-prediction depends on the position of the wind farm in the power grid. \blue{Indeed, the equilibrium regression systematically over-predicts power generation at bus $5$ and under-predicts power generation at buses $16$, $21$, and $23$. Interestingly, at bus $3$ the equilibrium suggests little to no corrections to the baseline forecast, and at bus $7$ the equilibrium suggests bias in different directions depending on the wind speed.}

We now focus on the driving forces behind such equilibrium predictions. For that, we first study the market revenues using two metrics: competitive ratio (CR) and revenue incentive to deviate $\Delta\mathcal{R}$. The CR is defined as a relative revenue performance with respect to the oracle and lies between $0$\% and $100$\%. The revenue incentive to deviate $\Delta\mathcal{R}$ is defined as the additional revenue gained from a unilateral deviation from the current regression model. It is computed by re-solving the NE problem \eqref{NE}--\eqref{comp_cond} for each wind power producer, having the regression models of all other wind power producers fixed, and then taking the difference. The results, summarized in Table \ref{tab:profits}, demonstrate increasing profits when wind power producers stick to their equilibrium models. Indeed, the CR increased for all wind farms, and for all but one it jumped from ``80s'' to ``90s'' in comparison with the baseline forecast. \blue{The revenues in Table \ref{tab:profits} demonstrate that the baseline regression is highly suboptimal for each wind power producer: any agent can increase their revenue by deviating from the baseline model while others stick to them. At equilibrium, the unilateral incentives to deviate from equilibrium regressions, on the training dataset, are marginal (due to a small weight parameter $\gamma$, used to regularize the equilibrium), asserting the result of Theorem \ref{th:ne_existence_uniqness} (inequality \eqref{eq_inequality})}. These incentives are a bit more than marginal on the unseen testing dataset, yet remain small compared to those in the baseline case. \blue{Last but not least, the demand charges under the baseline and equilibrium solutions increase relative to the oracle solution. This can be interpreted as the price of uncertainty paid by inflexible loads in the day-ahead market. Moreover, since the equilibrium solution tends to withhold zero-cost wind power generation from the day-ahead market (the aggregated forecast under-predicts the wind power generation), the equilibrium solution tends to be slightly more expensive for inelastic demands than the baseline case. This is consistent with prior work in \cite{morales2014electricity}.}

Table \ref{tab:system} summarizes the prediction and system outcomes under baseline and equilibrium regressions and contrasts them with the oracle. While the root mean square error (RMSE) of the equilibrium regression is notably higher, the equilibrium leads to lower dispatch costs than the baseline regression. This asserts Corollary \ref{corollary}, stating that the equilibrium achieves the \blue{regularized} least-cost dispatch across the two markets on average. Interestingly, the equilibrium yields larger day-ahead costs, as it tends to withhold zero-cost wind power generation from the day-ahead market. However, it also features real-time costs that are ten times smaller than in the baseline case. As a result, the average cost error in equilibrium is half as small as under the baseline regression. \blue{Another important benefit is that  the equilibrium regression leads to a {\it substantially} lower $\text{CVaR}_{5\%}$ estimation of dispatch cost, defined as the average cost across $5\%$ of the worst-case scenarios.} Importantly, the cost statistics are similar on training and testing datasets, highlighting the ability of the equilibrium regression to generalize beyond the training dataset.

We next study how the equilibrium drives regression model specification. Figure \ref{fig:feature} shows the optimal feature selection for the baseline regression, identical for all producers, and the equilibrium selection. Although all wind farms optimize on the same datasets, the equilibrium requires them to select features differently, depending on the wind farm's position in the power grid. \blue{This result highlights that the equilibrium regression features are more a function of LMPs than of the physical process underlying the data.}   

\begin{table*}
\caption{\blue{Average producer revenues \blue{($\mathcal{R}$)} and demand charges on the training(testing) dataset, \$}}
\vspace{-0.75em}
\label{tab:profits}
\centering
\begin{minipage}{1\linewidth}
\centering
\addtolength{\tabcolsep}{-0.3em}
\begin{tabular}{lrrrrrrrr}
\toprule
\multirow{2}{*}{Regression} & \multicolumn{6}{c}{Wind power producers} & \multicolumn{1}{c}{\multirow{2}{*}{\begin{tabular}[l]{@{}c@{}}Conventional \\ generators\end{tabular}}} & \multicolumn{1}{c}{\multirow{2}{*}{\begin{tabular}[l]{@{}c@{}}Demands\end{tabular}}} \\
\cmidrule(lr){2-7}
 & \multicolumn{1}{c}{bus $3$} & \multicolumn{1}{c}{bus $5$} & \multicolumn{1}{c}{bus $7$} & \multicolumn{1}{c}{bus $16$} & \multicolumn{1}{c}{bus $21$} & \multicolumn{1}{c}{bus $23$} & \multicolumn{1}{c}{} & \multicolumn{1}{c}{} \\
\midrule
Oracle & $756$ ($762$) & $795$ ($801$) & $696$ ($697$) & $720$ ($724$) & $507$ ($512$) & $700$ ($704$) & $23,464$ ($23,383$) & $32,210$ ($32,148$) \\   
\cmidrule(lr){1-9}
Baseline & $670$ ($677$) & $709$ ($717$) & $608$ ($611$) & $633$ ($638$) & $420$ ($426$) & $613$ ($619$) & $24,052$ ($23,952$) & $32,313$ ($32,237$) \\   
\multicolumn{1}{r}{CR} & $88.6$ ($88.8$) & $89.2$ ($89.5$) & $87.4$ ($87.7$) & $87.9$ ($88.1$) & $82.8$ ($83.2$) & $87.6$ ($87.9$) & $102.5$ ($102.4$) & $100.3$ ($100.3$) \\    
\multicolumn{1}{r}{$\Delta \mathcal{R}$} & $41$ ($37$) & $31$ ($29$) & $67$ ($58$) & $37$ ($34$) & $32$ ($30$) & $37$ ($34$) & \multicolumn{1}{c}{--------} & \multicolumn{1}{c}{--------} \\
\cmidrule(lr){1-9}
Equilibrium & $700$ ($706$) & $741$ ($754$) & $638$ ($644$) & $669$ ($668$) & $451$ ($450$) & $649$ ($646$) & $24,258$ ($24,155$) & $32,721$ ($32,627$) \\   
\multicolumn{1}{r}{CR} & $92.6$ ($92.7$) & $93.2$ ($94.1$) & $91.7$ ($92.4$) & $92.9$ ($92.3$) & $89$ ($87.9$) & $92.7$ ($91.8$) & $103.4$ ($103.3$) & $101.6$ ($101.5$) \\
\multicolumn{1}{r}{$\Delta \mathcal{R}$} & $-0$ ($-1$) & $-0$ ($-7$) & $-0$ ($-5$) & $0$ ($4$) & $0$ ($6$) & $0$ ($5$) & \multicolumn{1}{c}{--------} & \multicolumn{1}{c}{--------} \\
\bottomrule
\end{tabular}
\vspace{0.2em}

CR -- competitive ratio, i.e., relative performance with respect to oracle, \%\\
$\Delta \mathcal{R}$ -- average revenue incentive to unilaterally change a regression model, \$
\vspace{-1em}
\end{minipage}
\end{table*}

\begin{table*}
\caption{\blue{Forecast errors, dispatch cost, and cost errors on the training(testing) dataset}}
\vspace{-0.75em}
\label{tab:system}
\centering
\begin{tabular}{lrrrrrr}
\toprule
\multicolumn{1}{l}{\multirow{2}{*}{\begin{tabular}[l]{@{}l@{}}Regression\end{tabular}}} & \multicolumn{1}{c}{\multirow{2}{*}{RMSE, MWh}} & \multicolumn{3}{c}{Average dispatch cost, \$}                                                      & \multicolumn{2}{c}{Total dispatch cost error, \$}            \\
\cmidrule(lr){3-5}
\cmidrule(lr){6-7}
\multicolumn{1}{c}{}                          & \multicolumn{1}{c}{}                           & \multicolumn{1}{c}{total} & \multicolumn{1}{c}{day-ahead} & \multicolumn{1}{c}{real-time} & \multicolumn{1}{c}{average} & \multicolumn{1}{c}{$\text{CVaR}_{5\%}$} \\
\midrule
Oracle & \multicolumn{1}{c}{--------} & $21,102$ $(21,046)$ & $21,102$ $(21,046)$ & \multicolumn{1}{c}{--------} & \multicolumn{1}{c}{--------} & \multicolumn{1}{c}{--------} \\  
Baseline & $46$ $(46)$ & $21,580$ $(21,520)$ & $21,136$ $(21,092)$ & $444$ $(428)$ & $478$ $(474)$ & $2,375$ $(2,402)$ \\ 
Equilibrium & $53$ $(54)$ & $21,433$ $(21,379)$ & $21,499$ $(21,456)$ & $-66$ $(-77)$ & $331$ $(333)$ & $1,245$ $(1,246)$ \\ 
\bottomrule
&\end{tabular}
\end{table*}
\begin{figure}
    \centering
    \includegraphics[width=0.5\textwidth]{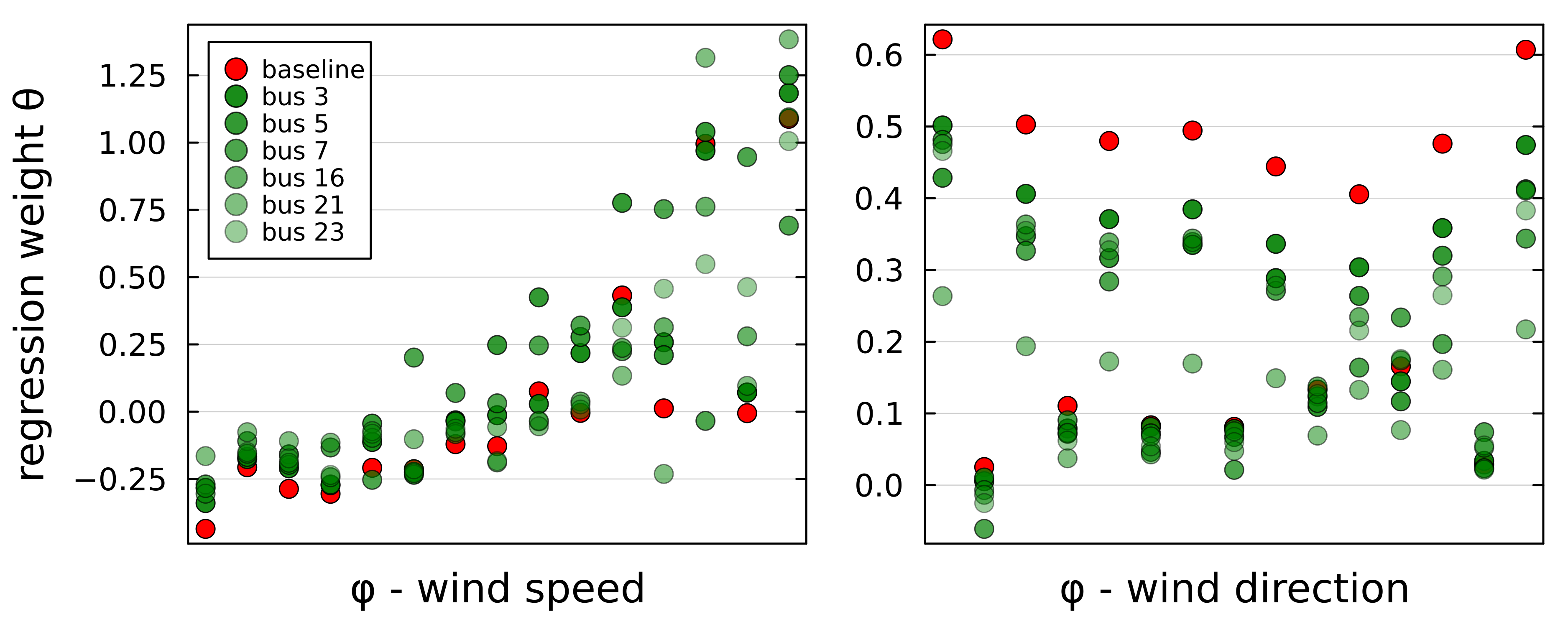}
    \caption{\blue{The regression weights assigned to each of the 30 kernel feature transformations (15 kernels for wind speed and 15 kernels for wind direction) under the baseline method (red) and in equilibrium (green).}}
    \label{fig:feature}
\end{figure}

We conclude by discussing computational aspects. The results in this section are obtained by solving a centralized optimization problem from Corollary \ref{corollary}, which is solved using Mosek's interior-point solver \cite{aps2020mosek}, requiring less than 80s to build and converge on the standard laptop. The proposed ADMM algorithm, with the following schedule for the step size $\varrho=100_{(k\leqslant199)}, 10_{(k\geqslant200)}, 5_{(k\geqslant230)}, 1_{(k\geqslant275)}$, yields identical results. Each iteration takes 129.5s on average, and the whole algorithm runs 10.8h. The ADMM algorithm, however, is amenable to further improvements such as parallelization and adaptive step size selection \cite{boyd2011distributed}.

\section{Conclusions}

\blue{In this paper, we established the concept of regression equilibrium, where wind power producers optimize their forecast models against prices in day-ahead and real-time markets. This not only maximizes their profits but also implicitly improves the coordination between these markets, ensuring the least-cost dispatch solution.} The results on the standard IEEE test system demonstrated that equilibrium leads to significant improvements in wind power profits as well as a reduction in dispatch costs, with more significant improvements in the worst-case tail of the cost distribution. Moreover, we found that the system-optimal forecast features vary among wind power producers, depending on their location in the grid.

\blue{This work points to the new research direction, where ML models are reconsidered in terms of their market impacts and the ability of existing markets to sustain potentially drastic impacts that these models may have.} \blue{There are several avenues for future work. First, we plan to include more ML users and relax the energy-only market assumption. For example, we will model self-committing generation units deciding their commitment status using trained classifiers in the spirit of \cite{10530167}. This way, we will study market interactions among heterogeneous ML models. Next, we will include the virtual bidders \cite{hogan2016virtual} who are likely to be affected by the new regression equilibrium. Specifically, as the regression equilibrium improves the temporal coordination between the day-ahead and real-time markets, the virtual bidders will have less inefficiency to profit from and are likely to see smaller profit margins.}

\appendix
\subsection{Proof of Theorem \ref{th:ne_existence_uniqness}.}\label{app:one}
{\it Existence.} Feasible regions $\mathcal{O},$ $\mathcal{G}$ and $\mathcal{S}$ are convex, and their intersection with a convex set of equilibrium constraints $\mathcal{E}$ is compact per Assumption \ref{ass:5}. Moreover, the mapping $F$ is continuous as profit functions in \eqref{eq:revenues} are all differentiable. Hence, by Corollary 2.2.5 of the existence of the solution to the VI from \cite{facchinei2003finite}, the equilibrium solution exists. 

{\it Uniqueness.} We first show the existence of an equivalent convex optimization problem that solves the VI, and then show the uniqueness through the properties of this optimization. Towards finding the equivalent optimization, we observe that the Jacobian matrix of $F(\mbg{\Theta},\mb{G},\mb{L})$, i.e.,
{\setlength{\arraycolsep}{1pt}
\begin{align}\label{hassian}
    &\!\!\!\nabla F(\mbg{\Theta},\mb{G},\mb{L}) =\nonumber\\
    &\!\!\!=\frac{2}{n}
     \begin{bmatrix*}[r]
        \mbg{\Gamma}\!\smallotimes\!\sum_{i=1}^{n}\mbg{\varphi}_{i}\mbg{\varphi}_{i}^{\top}\!\!\! & \mb{0} & \mb{0} & \mb{0} & \mb{0} \\
        \mb{0}&\mb{I}_{n}\!\smallotimes\!\mb{C} & \mb{I}_{n}\!\smallotimes\!\mb{C}  & -\mb{I}_{n}\!\smallotimes\!\mb{C} & \mb{0}\\
        \mb{0}&\mb{I}_{n}\!\smallotimes\!\mb{C} & \mb{I}_{n}\!\smallotimes\!\mb{C} & -\mb{I}_{n}\!\smallotimes\!\mb{C} & \mb{0}\\
        \mb{0}&-\mb{I}_{n}\!\smallotimes\!\mb{C} & -\mb{I}_{n}\!\smallotimes\!\mb{C} & \mb{I}_{n}\!\smallotimes\!\mb{C} & \mb{0} \\
        \mb{0}& \mb{0} & \mb{0} & \mb{0} & \mb{I}_{n}\!\smallotimes\!\mb{S}
     \end{bmatrix*}
\end{align}}
\hspace{-3pt}is a symmetric matrix. By the Symmetry Principle Theorem 1.3.1 in \cite{facchinei2003finite}, this Jacobian matrix renders function $F$ a gradient map, meaning that there exists a scalar function $f(\mb{x})$, such that $F(\mb{x})=\nabla f(\mb{x})$ for all $\mb{x}\in\mathcal{K}$. Function $f$ can be found by integrating the gradient map as $f(\mb{x}) =\int_{0}^{1}F(\mb{x}_{0}+t(\mb{x}-\mb{x}_{0}))^{\top}(\mb{x}-\mb{x}_{0})dt$ \cite[\S1]{facchinei2003finite}. In our case, the integration of the gradient map $F\left(\mbg{\Theta},\mb{G},\mb{L}\right)$ takes the form:
\begin{align}
    &f(\mbg{\Theta},\mb{G},\mb{L})=\nonumber\\
    =& \int_{0}^{1}
    \frac{1}{n}\begin{bmatrix*}[l]
        \\[-0.75em]
        2\textstyle\sum_{i=1}^{n}\gamma_{1}((t\cdot\mbg{\theta}_{1})^{\top}\mbg{\varphi}_{i}-w_{1i})\cdot\mbg{\varphi}_{i}\\[-0.25em]
        \hspace{0.7em}\dots\\[-0.05em]
        2\textstyle\sum_{i=1}^{n}\gamma_{b}((t\cdot\mbg{\theta}_{b})^{\top}\mbg{\varphi}_{b}-w_{bi})\cdot\mbg{\varphi}_{i}\\[0.5em]
        \mb{c}+2\mb{C}(\mb{p}_{1}+\mb{r}_{1}^{+}-\mb{r}_{1}^{-})\cdot t \\[-0.25em]
        \hspace{0.7em}\dots\\[-0.05em]
        \mb{c}+2\mb{C}(\mb{p}_{n}+\mb{r}_{n}^{+}-\mb{r}_{n}^{-})\cdot t\\[0.5em]
        \mb{c}^{+}+2\mb{C}(\mb{p}_{1}+\mb{r}_{1}^{+}-\mb{r}_{1}^{-})\cdot t \\[-0.25em]
        \hspace{0.7em}\dots\\[-0.05em]
        \mb{c}^{+}+2\mb{C}(\mb{p}_{n}+\mb{r}_{n}^{+}-\mb{r}_{n}^{-})\cdot t\\[0.5em]
        -\mb{c}^{-}-2\mb{C}(\mb{p}_{1}+\mb{r}_{1}^{+}-\mb{r}_{1}^{-})\cdot t \\[-0.25em]
        \hspace{0.7em}\dots\\[-0.05em]
        -\mb{c}^{-}-2\mb{C}(\mb{p}_{n}+\mb{r}_{n}^{+}-\mb{r}_{n}^{-})\cdot t\\[0.5em]
        \mb{s}+2\mb{S}\mbg{\ell}_{1}\cdot t \\[-0.25em]
        \hspace{0.7em}\dots\\[-0.05em]
        \mb{s}+2\mb{S}\mbg{\ell}_{n}\cdot t\\[0.4em]
    \end{bmatrix*}^{\top}
    \begin{bmatrix*}[l]
    \\[-0.95em]
    \mbg{\theta}_{1}\\[-0.5em]
    \hspace{0.2em}\vdots\\[-0.15em]
    \mbg{\theta}_{b}\\
    \mb{p}_{1}\\[-0.5em]
    \hspace{0.2em}\vdots\\[-0.5em]
    \mb{p}_{n}\\[0.1em]
    \mb{r}_{1}^{+}\\[-0.5em]
    \hspace{0.2em}\vdots\\[-0.5em]
    \mb{r}_{n}^{+}\\[0.1em]
    \mb{r}_{1}^{-}\\[-0.5em]
    \hspace{0.2em}\vdots\\[-0.5em]
    \mb{r}_{n}^{-}\\[0.1em]
    \mbg{\ell}_{1}\\[-0.5em]
    \hspace{0.2em}\vdots\\[-0.35em]
    \mbg{\ell}_{n}\\
    \end{bmatrix*}dt = \nonumber\\
    =&
    \frac{1}{n}\left(\!
    2\sum_{i=1}^{n}\gamma_{1}\left(\!(\mbg{\theta}_{1}^{\top}\mbg{\varphi}_{i})^{\top} (\mbg{\theta}_{1}^{\top}\mbg{\varphi}_{i})\int_{0}^{1}tdt - w_{1i}\mbg{\theta}_{1}^{\top}\mbg{\varphi}_{i}\!\right)
    \!\!\right)\nonumber\\[-0.25em]
    &\dots\nonumber\\[-0.25em]
    &
    +\frac{1}{n}\left(\!
    2\sum_{i=1}^{n}\gamma_{b}\!\left(\!\!(\mbg{\theta}_{b}^{\top}\mbg{\varphi}_{i})^{\top} (\mbg{\theta}_{b}^{\top}\mbg{\varphi}_{i})\int_{0}^{1}tdt - w_{1i}\mbg{\theta}_{b}^{\top}\mbg{\varphi}_{i}\!\right)
    \!\!\right)\nonumber\\
    &+\frac{1}{n}\Big(\sum_{i=1}^{n}2(\mb{p}_{i}+\mb{r}_{i}^{+}-\mb{r}_{i}^{-})\mb{C}(\mb{p}_{i}+\mb{r}_{i}^{+}-\mb{r}_{i}^{-})\int_{0}^{1}tdt\nonumber\\
    &\quad\quad\quad+\mb{c}^{\top}\mb{p}_{i}+\mb{c}^{+\top}\mb{r}_{i}^{+}-\mb{c}^{-\top}\mb{r}_{i}^{-}\Big)\nonumber\\
    &+\frac{1}{n}\Big(\sum_{i=1}^{n}2\mbg{\ell}_{i}^{\top}\mb{S}\mbg{\ell}_{i}\int_{0}^{1}tdt+\mb{s}^{\top}\mbg{\ell}_{i}\Big)\nonumber\\
    =&
    \frac{1}{n}\left(
    \sum_{i=1}^{n}\gamma_{1}\left((\mbg{\theta}_{1}^{\top}\mbg{\varphi}_{i}-w_{1i})^2 + w_{1i}^2\right)\right)\hfill\textcolor{gray}{\text{\scriptsize// by completing the square}}\nonumber\\[-0.25em]
    &\dots\nonumber\\[-0.25em]
    &
    +\frac{1}{n}\left(
    \sum_{i=1}^{n}\gamma_{b}\left((\mbg{\theta}_{b}^{\top}\mbg{\varphi}_{i}-w_{bi})^2 + w_{bi}^2\right)\right)\nonumber\\
    &+\frac{1}{n}\Big(\sum_{i=1}^{n}(\mb{p}_{i}+\mb{r}_{i}^{+}-\mb{r}_{i}^{-})\mb{C}(\mb{p}_{i}+\mb{r}_{i}^{+}-\mb{r}_{i}^{-})\nonumber\\
    &\quad\quad\quad+\mb{c}^{\top}\mb{p}_{i}+\mb{c}^{+\top}\mb{r}_{i}^{+}-\mb{c}^{-\top}\mb{r}_{i}^{-}\Big)\nonumber\\
    &+\frac{1}{n}\Big(\sum_{i=1}^{n}\mbg{\ell}_{i}^{\top}\mb{S}\mbg{\ell}_{i}+\mb{s}^{\top}\mbg{\ell}_{i}\Big)\nonumber\\
    =&
    \frac{1}{n}\sum_{i=1}^{n}\Big(\norm{\mbg{\Theta}\mbg{\varphi}_{i}-\mb{w}_{i}}_{\mbg{\Gamma}}^{2} + \mb{w}_{i}^{\top}\mbg{\Gamma}\mb{w}_{i}\nonumber\\[-0.75em]
    &\quad\quad\quad\;\;+\norm{\mb{p}_{i}+\mb{r}_{i}^{+}-\mb{r}_{i}^{-}}_{\mb{C}}^{2}+\mb{c}^{\top}\mb{p}_{i}\nonumber\\
    &\quad\quad\quad\quad\quad\;\;+\mb{c}^{+\top}\mb{r}_{i}^{+}-\mb{c}^{-\top}\mb{r}_{i}^{-} + \norm{\mbg{\ell}_{i}}_\mb{S}^{2}+\mb{s}^{\top}\mbg{\ell}_{i}\Big),\label{eq:fun_f}
\end{align}
which is a convex function of NE variables. Note, w.l.o.g., we used $\mb{x}_{0}=\mb{0}$. For non-zero $\mb{x}_{0}$, function $f$ would include additional terms independent from $\mbg{\Theta},\mb{G},$ and $\mb{L}$. The virtue of the symmetry principle is that our $\text{VI}(\mathcal{K},F)$ becomes a stationary point problem of the following optimization problem:
\begin{subequations}\label{eqopt}
\begin{align}
    \minimize{\mbg{\Theta},\mb{G},\mb{L}}\quad&f(\mbg{\Theta},\mb{G},\mb{L})\\
    \st\quad&\mbg{\Theta},\mb{G},\mb{L}\in\mathcal{K},
\end{align}
\end{subequations}
such that solving problem \eqref{eqopt} solves the VI. Function $f$ is strictly convex, because \blue{its Hessian $\nabla F$ in \eqref{hassian} is positive-definite. Indeed, cost parameters $\mb{C}\in\mathbb{S}_{+}$, $\mb{S}\in\mathbb{S}_{+}$ are positive definite matrices. The transformed feature vector $\mbg{\varphi}$ is always positive due the Gaussian kernel functions forming $\mbg{\varphi}$, and matrix $\mbg{\Gamma}$ is diagonal with positive diagonal entries; thus, the product $\mbg{\Gamma}\!\smallotimes\!\sum_{i=1}^{n}\mbg{\varphi}_{i}\mbg{\varphi}_{i}^{\top}$ is also positive-definite}. The solution to problem \eqref{eqopt}---and hence to the VI---is thus unique, and inequality \eqref{eq_inequality} underlying the equilibrium holds. 

We finally observe that $\mb{w}_{i}^{\top}\mbg{\Gamma}\mb{w}_{i}$ in \eqref{eq:fun_f} is constant and can be disregarded in optimization, so minimizing $f(\mbg{\Theta},\mb{G},\mb{L})$ is equivalent to minimizing the average social cost of electricity, regularized by private regression losses. Hence, the equilibrium solution aligns with the regularized social optimum. 

\balance
\bibliography{references.bib}
\bibliographystyle{IEEEtran}

\endgroup
\end{document}